\documentclass[twocolumn,aps,superscriptaddress,floatfix]{revtex4}
\usepackage{amsmath}
\usepackage{amssymb}
\usepackage{graphicx}
\usepackage{color}

\newcommand{\bea}{\begin{eqnarray}}
\newcommand{\eea}{\end{eqnarray}}
\newcommand{\bse}{\begin{subequations}}
\newcommand{\ese}{\end{subequations}}

\newcommand{\emb}{EuMg$_2$Bi$_2$}
\newcommand{\cas}{CaAl$_2$Si$_2$}
\newcommand{\ymb}{YbMg$_2$Bi$_2$}

\begin{document}

\title{Magnetic, thermal, and electronic-transport properties of EuMg$_2$Bi$_2$ single crystals}

\author{Santanu Pakhira}
\affiliation{Ames Laboratory, Ames, Iowa 50011, USA}
\author{M. A. Tanatar}
\affiliation{Ames Laboratory, Ames, Iowa 50011, USA}
\affiliation{Department of Physics and Astronomy, Iowa State University, Ames, Iowa 50011, USA}
\author{D. C. Johnston}
\affiliation{Ames Laboratory, Ames, Iowa 50011, USA}
\affiliation{Department of Physics and Astronomy, Iowa State University, Ames, Iowa 50011, USA}

\date{\today}

\begin{abstract}

The trigonal compound \emb\ has recently been discussed in terms of its topological band properties.  These are intertwined with its magnetic properties.  Here detailed studies of the magnetic, thermal, and electronic transport properties of \emb\ single crystals are presented.  The Eu$^{+2}$ spins-7/2 in \emb\ exhibit an antiferromagnetic (AFM) transition at a temperature $T_{\rm N} =  6.7$~K, as previously reported.  By analyzing the anisotropic magnetic susceptibility $\chi$ data below $T_{\rm N}$ in terms of molecular-field theory (MFT), the AFM structure is inferred to be a $c$-axis helix, where the ordered moments in the hexagonal $ab$-plane layers are aligned ferromagnetically in the $ab$ plane with a turn angle between the moments in adjacent moment planes along the $c$~axis of $\approx 120^\circ$.  An alternate but less likely magnetic structure is a planar structure with nearest-neighbor Eu spins aligned at $\approx 120^\circ$ with respect to each other, where these ordered-moment layers are stacked along the $c$~axis.  The magnetic heat capacity exhibits a $\lambda$ anomaly at $T_{\rm N}$ with evidence of dynamic short-range magnetic fluctuations both above and below $T_{\rm N}$.  The high-$T$ limit of the magnetic entropy is close to the theoretical value for spins-7/2.   The in-plane electrical resistivity $\rho(T)$ data indicate metallic character with a mild and disorder-sensitive upturn below $T_{\rm min}=23$~K\@.  An anomalous rapid drop in $\rho(T)$ on cooling below $T_{\rm N}$ as found in zero field is replaced by a two-step decrease in magnetic fields. The $\rho(T)$ measurements also reveal an additional transition below $T_{\rm N}$ in applied fields of unknown origin that is not observed in the other measurements and may be associated with an incommensurate to commensurate AFM transition.  The dependence of $T_{\rm N}$ on the $c$-axis magnetic field $H_\perp$ was derived from the field-dependent $\chi(T)$, $C_{\rm p}(T)$, and $\rho(T)$ measurements.  This $T_{\rm N}(H_\perp)$ was found to be consistent with the prediction of MFT for a $c$-axis helix with $S=7/2$ and was used to generate a phase diagram in the $H_\perp$-$T$ plane.

\end{abstract}

\maketitle

\section{Introduction}

The interplay between magnetism and band topology has generated immense interest recently due to the discovery of nontrivial phenomena such as the quantum anomalous Hall effect (QAHE)~\cite{Qi2006_QAHE, Liu2008_QAHE, Yu2010_QAHE, Chang2013_QAHE, Liu2016_QAHE}, axion electrodynamics~\cite{Essin2009_AED, Li2010_AED, Lee_AED, Wu2016_AED}, and realization of relativistic particles like Majorana fermions~\cite{Do2017_Majorana, Akhmerov2009_Majorana, Cook2011_Majorana, Xu2015_Majorana}. A topological state manifests topologically-protected electronic surface states that are different in nature from the bulk states~\cite{Fu2007_TI, Hasan2010_TI, Moore2010_TI, Qi2011_TI,  Ando2013_TI, Tokura2019_TI}.  In addition to the parabolic bulk band, depending upon symmetry preservation, a unique gapless/gapped surface state with a linear energy versus crystal momentum $E$-$k$ relation is also observed in these materials, resulting in massless surface electrons with ultrahigh mobility.

Time reversal symmetry (TRS) invariance in topological insulating states and its breaking in the presence of magnetism play a key role in discoveries in this field. Exotic quantum phenomena such as QAHE, chiral Majorana modes, and the topological magnetoelectric effect (TME) have been predicted in topological materials based on TRS breaking by magnetic order, with experimental support~\cite{Liu2008_QAHE, Yu2010_QAHE, Chang2013_QAHE, Essin2009_AED, Li2010_AED, Do2017_Majorana, Fu2007_TI, Qi2008_TRB, Garate2010_TRB, Xu2012_TRB}. In a topological material exhibiting the QAHE, electrons can carry a dissipationless current, which is thus promising for use in energy-efficient electronic devices and also for fast computing. The experimental realization of QAHE in magnetic topological insulators has paved the way for researchers to discover other novel phenomena by studying various magnetic topological systems. In order to fully understand the phenomena triggered by TRS breaking in such magnetic topological systems, it is essential to understand the magnetic structure and symmetry of those materials.

Rare-earth-based intermetallic compounds have been of significant interest for many years due to their complex properties such as  superconductivity, heavy fermion behavior, valence fluctuations, giant magnetocaloric effect, Kondo behavior, and quantum criticality~\cite{Fertig1977_IC, Nagarajan1994_IC, Ghosh1995_IC, Curro2000_IC, Hundley2001_IC, Sakai2011_IC, Yamaoka2014_IC, Pecharsky1997_IC, Pakhira2016_IC, Pakhira2017_IC, Buschow1969_IC, Gignoux1984_IC, Ishida2002_IC, Arndt2011_IC}. Several rare-earth-based intermetallic compounds have also been discovered recently to exhibit topological states coupled with magnetic interactions~\cite{Hirschberger2016_GdPtBi, Wang2016_YbMnBi2,May2014_EuMnBi2,Soh2019_EuMnSb2,Schellenberg2011_EuCd2As2,Jo2020_EuCd2As2,Xu2019_EuIn2As2}. These discoveries reveal the subtle importance of magnetism in controlling the electronic surface states in the magnetic topological materials.

Recently, multiple Dirac states at different energies with respect to the Fermi energy were reported in a new magnetic topological material \emb~\cite{Kabir2019}. The magnetic ground state of the compound was suggested to be AFM in nature below $T_{\rm N} \approx 7$~K, the details of which are unclear as yet~\cite{Kabir2019, May2011, Ramirez2015}. Based on the magnetic susceptibility behavior, where the $c$-axis susceptibility $\chi_c$ decreases somewhat and the $ab$-plane susceptibility $\chi_{ab}$ is almost temperature independent below $T_{\rm N}$, it was suggested that the moments are aligned along the $c$~axis~\cite{Kabir2019}, presumably in a collinear AFM structure. In order to understand the mechanism of topological surface states and also to tune its nature it is important to clarify the magnetic structure and its evolution with applied magnetic field in EuMg$_2$Bi$_2$.

\begin{figure}
\includegraphics [width=2.25in]{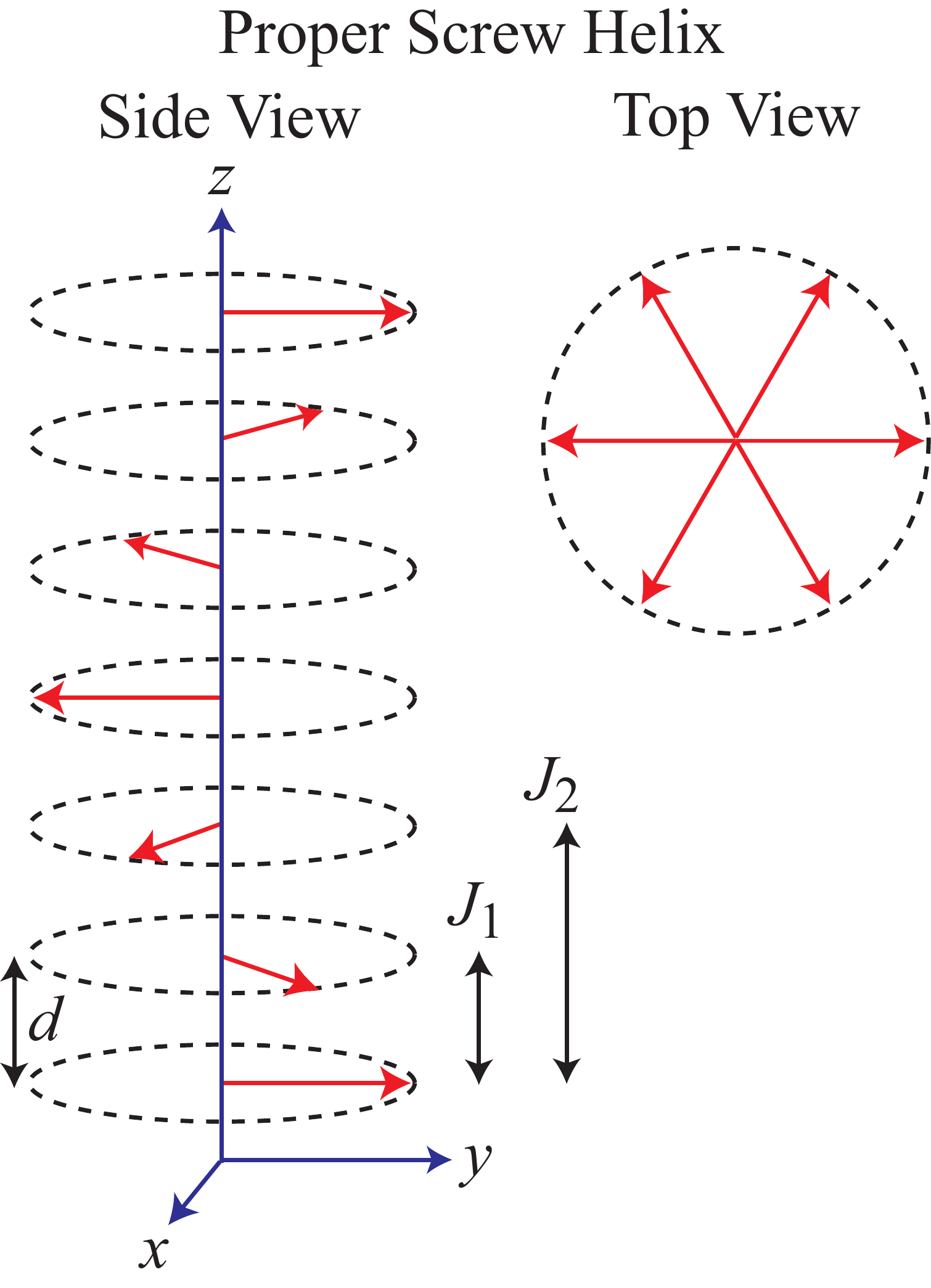}
\caption {Generic helical AFM structure~\cite{Johnston2012, Johnston2015}.  Each arrow represents a layer of moments perpendicular to the $z$~axis that are ferromagnetically aligned within the $xy$ plane and with interlayer separation~$d$.  The wave vector {\bf k} of the helix is directed along the $z$~axis.  The magnetic moment turn angle between adjacent magnetic layers is $kd$.  The exchange interactions $J_{1}$ and $J_{2}$ within the one-dimensional $J_0$-$J_{1}$-$J_{2}$ Heisenberg MFT model are indicated.}
\label{Fig:Fig_Helix}
\end{figure}

\begin{figure}
\includegraphics [width=1.75in]{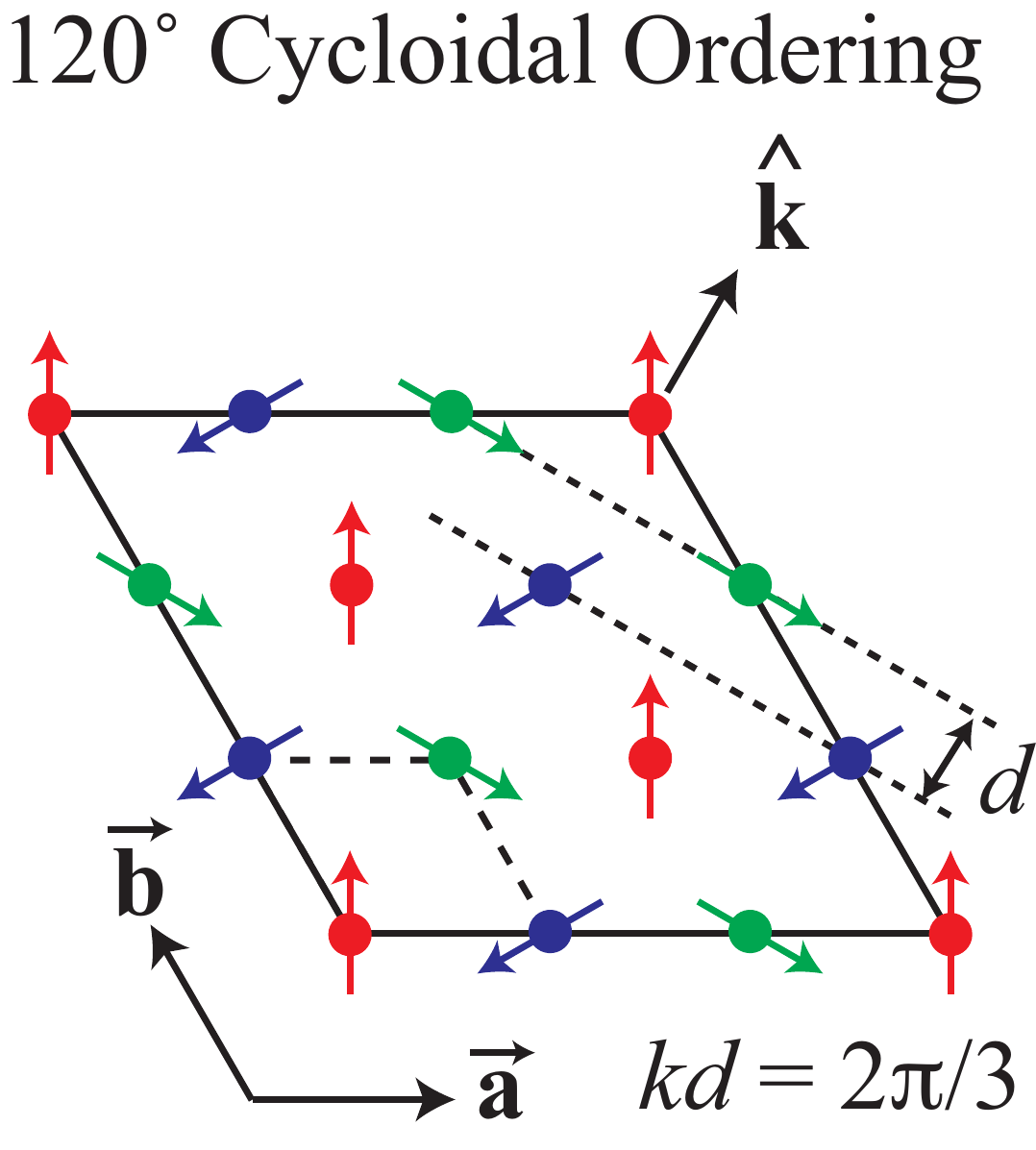}
\caption{Schematic illustration of a 120$^\circ$ cycloidal helix AFM structure on a triangular lattice that occurs when the wave vector {\bf k} of a cycloid is in the $xy$-plane in which the ordered magnetic moments reside. The definitions of $d$ and the exchange constants are the same as in Fig.~\ref{Fig:Fig_Helix}.}
\label{Fig:Fig120degCycloid}
\end{figure}

 The primary purpose of this work was to investigate the magnetic structure and to explore the magnetic phase diagram of \emb\ as a function of temperature and applied magnetic field $H$ to provide an experimental basis for relating the properties to the topology of the theoretical band structure.  We present detailed measurements of the anisotropic $\chi$ versus temperature~$T$ and magnetization $M$ versus $H$ isotherms for \emb\ single crystals, together with complementary heat-capacity and electrical-resistivity measurements.  We find that the magnetic susceptibilities for fields applied both in-plane and out-of-plane are almost temperature independent below $T_{\rm N}$, which is very unusual in magnetic Eu$^{+2}$-based systems.  Using a recently-developed molecular-field theory (MFT)~\cite{Johnston2012, Johnston2015}, we deduce that the nearly $T$-independent $\chi(T)$ data below $T_{\rm N}$ indicate that the magnetic structure is either a $c$-axis helix in which the Eu moments are ferromagnetically aligned within each $ab$~plane layer, with a turn angle $kd\approx120^\circ$ between the moments in adjacent layers as generically illustrated in Fig.~\ref{Fig:Fig_Helix}~\cite{Johnston2012, Johnston2015}, or a stacked coplanar structure with a $120^\circ$ angle between adjacent Eu moments in each triangular-lattice $ab$-plane layer as in the 120$^\circ$ cycloidal structure in Fig.~\ref{Fig:Fig120degCycloid}.  The in-plane electrical resistivity $\rho(T)$ data indicate metallic character with a mild and disorder-sensitive upturn below $T_{\rm min}=23$~K\@. An anomalous rapid drop on cooling below $T_{\rm N}$ as found in zero field is replaced by a two-step decrease in magnetic fields. The $\rho(T)$ measurements also reveal an additional transition below $T_{\rm N}$ in applied fields of unknown origin that is not observed in the other measurements and may be associated with an incommensurate to commensurate AFM transition.

The experimental details and the crystal structure studies are discussed in Sec.~\ref{Sec:ExpDet}.  The $\chi(T)$ and $M(H)$ isotherm data and analyses are given in Sec.~\ref{Sec:Magnetism}.  Our heat capacity $C_{\rm p}(H,T)$ data and analyses are presented in Sec.~\ref{Sec:Cp}, and the electrical resistivity $\rho(H,T)$ data and analyses in Sec.~\ref{Sec:rho}.  The paper concludes with a summary in Sec.~\ref{Sec:Summary}.

\section{\label{Sec:ExpDet} Experimental Details and Crystal Structure}

Single crystals of \emb\ and \ymb\ were grown with the flux method in two different ways using high purity elements Eu (Ames Laboratory), Mg (Alfa Aesar, 99.98\%), and Bi (Alfa Aesar, 99.9999\%). One batch of crystals was obtained from the nominal flux composition EuMg$_4$Bi$_6$ and YbMg$_4$Bi$_6$ as described elsewhere~\cite{May2011}. Another batch of crystals was grown from the starting composition EuMg$_2$Bi$_7$. The starting elements were loaded into alumina crucibles and sealed in silica tubes under $\approx 1/4$~atm of Ar gas.  A sealed tube was heated to 900~$^{\circ}$C at a rate of 50~$^{\circ}$C/h and held at that temperature for 12 h. Then it was cooled to 500~$^{\circ}$C over 200 h and the single crystals were then obtained by decanting the excess flux using a centrifuge.  Most of the crystals obtained are three dimensional in shape with clear trigonal facets.  In this work measurements were carried out on crystals obtained using the first growth process.

Room-temperature x-ray diffraction (XRD) measurements were carried out using a Rigaku Geigerflex x-ray diffractometer using \mbox{Cu-K$_\alpha$} radiation. Structural analysis was performed by Rietveld refinement using the Fullprof software package~\cite{Carvajal1993}. The chemical composition and  homogeneity of the crystals were checked using a JEOL scanning-electron microscope (SEM) equipped with an EDS (energy-dispersive x-ray spectroscopy) \mbox{analyzer.}

The $T$- and $H$-dependent magnetization measurements were performed using a magnetic-properties measurement system (MPMS, Quantum Design, Inc.) in the $T$ range 1.8--300~K and $H$ up to 5.5~T (1\,T~$\equiv10^4$\,Oe). A Quantm Design, Inc., physical-properties measurement system (PPMS) was used to measure $C_{\rm p}(H,T)$ and $\rho(H,T)$.  Contacts to the samples were soldered with In in standard four-probe configuration, similar to the technique used for FeSe \cite{FeSeresistivity}. The contact resistance was in the sub-Ohm range.

\begin{figure}
\includegraphics[width = 3.4in]{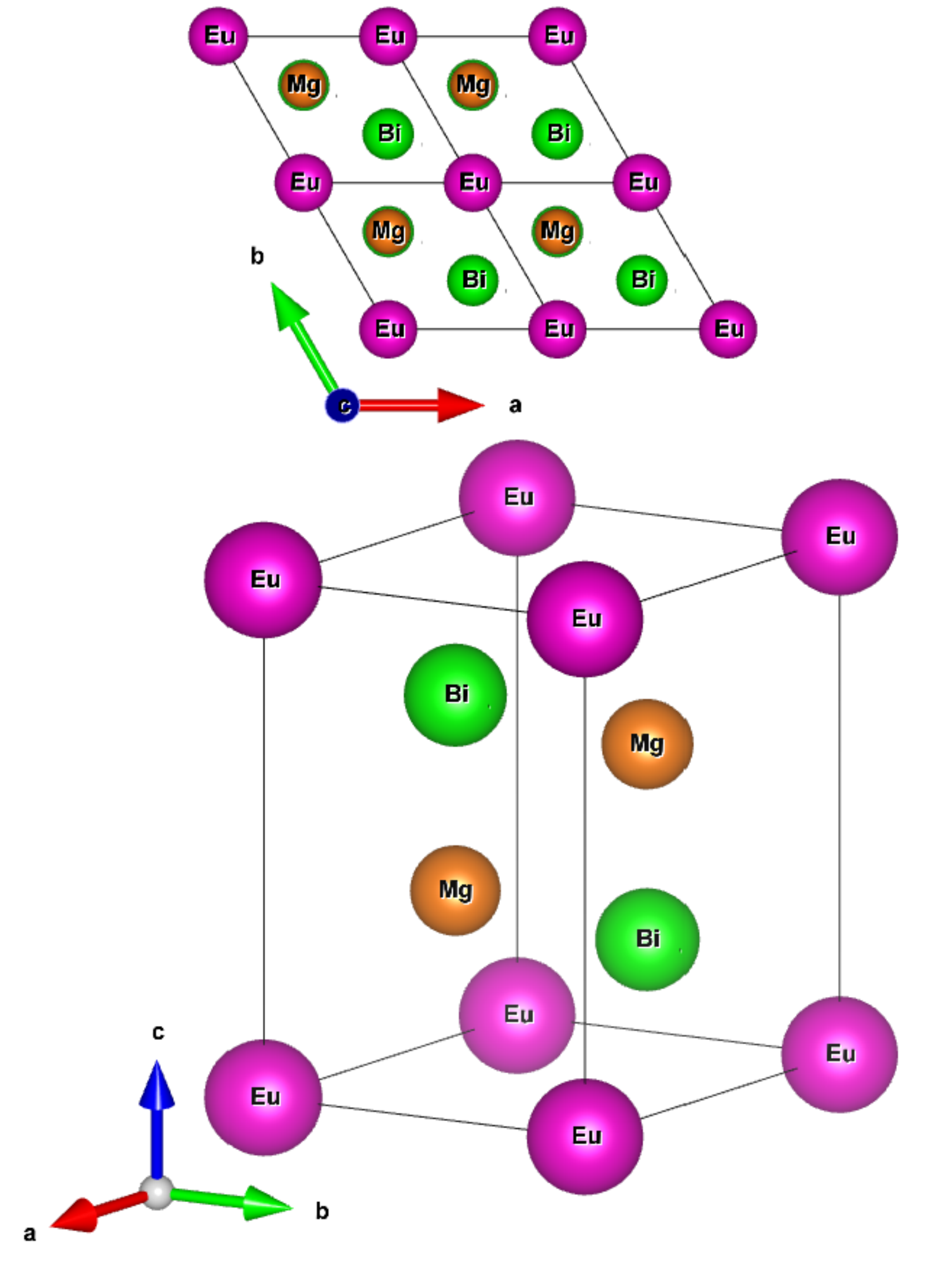}
\caption{EuMg$_2$Bi$_2$ crystal structure. (top)~Projection of the structure onto the hexagonal $ab$ plane.  (bottom)~Three-dimensional hexagonal unit cell.}
\label{EuMg2Bi2_structure}
\end{figure}

\begin{figure}
\includegraphics[width = 3.4in]{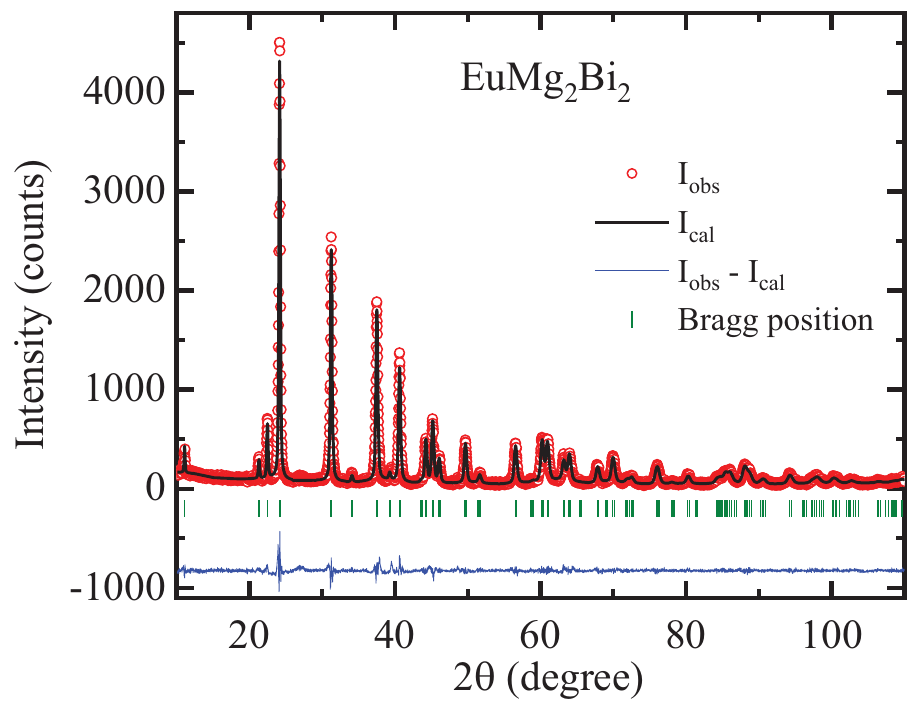}
\caption{Room temperature powder x-ray diffraction pattern of \emb . The solid black line through the red circle experimental data points is the Rietveld refinement calculated for the CaAl$_2$Si$_2$-type crystal structure (space group $P\bar{3}m1$). The green vertical bars are the allowed Bragg positions and the blue solid curve at the bottom represents the difference between the experimental and calculated diffraction patterns.}
\label{XRD}
\end{figure}

The reported \cas-type trigonal crystal structure of \emb\ (space group $P\bar{3}m1$)~\cite{May2011,Zheng1986} is shown in Fig.~\ref{EuMg2Bi2_structure}.  The structure consists of stacked planar triangular lattices of Eu in the $ab$ plane separated along the $c$~axis by two ordered MgBi layers.  The room-temperature powder x-ray diffraction (XRD) pattern collected on crushed \emb\ single crystals along with our Rietveld refinement are shown in Fig.~\ref{XRD}. The refinement confirms that \emb\ crystallizes in the \cas-type crystal structure with space group $P\bar{3}m1$. The refined parameters are summarized in Table~\ref{TableXRD}. The lattice parameters obtained are in agreement with previously reported values~\cite{May2011, Ramirez2015, Kabir2019}. The SEM-EDS measurements confirm the homogeneity of the grown crystals with an average composition EuMg$_{1.98(2)}$Bi$_{2.01(2)}$, which agrees with the stoichiometric composition~\emb\ to within the errors.

\begin{table}
\caption{\label{TableXRD} Crystallographic and refinement parameters obtained from the structural analysis of room temperature powder x-ray diffraction data of trigonal \emb\ (space group $P\bar{3}m1$).}
 \begin{center}
  \begin{tabular}{ c c c c c c  }
  \hline\hline
  \multicolumn{2}{c}{Hexagonal lattice parameters} & \multicolumn{4}{c}{}  \\
  \hline
  \multicolumn{2}{c}{{\it a}({\AA})} & \multicolumn{4}{c}{4.7724(3)}  \\
  \multicolumn{2}{c}{{\it c}({\AA})} & \multicolumn{4}{c}{7.8483(5)}  \\
  \multicolumn{2}{c}{{\it c}/{\it a}} & \multicolumn{4}{c}{1.644(1)}  \\
  \multicolumn{2}{c}{V$_{\rm cell}$ ({\AA}$^3$)} & \multicolumn{4}{c}{154.80(2)}  \\
   \multicolumn{2}{c}{Refinement quality} & \multicolumn{4}{c}{}  \\
   \hline
   \multicolumn{2}{c}{$\chi^2$} & \multicolumn{4}{c}{2.10}  \\
   \multicolumn{2}{c}{R$_{\rm p}$ (\%)} & \multicolumn{4}{c}{8.43}  \\
   \multicolumn{2}{c}{R$_{\rm {wp}}$ (\%)} & \multicolumn{4}{c}{11.5}  \\
       \multicolumn{2}{c}{Atomic coordinates} & \multicolumn{4}{c}{} \\
       \hline
  Atom & Wyckoff Symbol & {\it x} & {\it y} & {\it z} & occupancy (\%) \\

  Eu & 1\textit{a} & 0 & 0 & 0 & 100 \\
  Mg & 2\textit{d} & 1/3 & 2/3 & 0.6285(5) & 99(2) \\
  Bi & 2\textit{d} & 1/3 & 2/3 & 0.2499(4) & 98(2) \\
  \hline\hline
 \end{tabular}
\end{center}
\end{table}

\section{\label{Sec:Magnetism} Magnetic susceptibility versus temperature and magnetization versus field isotherms}

\subsection{Magnetic susceptibility}

\begin{figure}
\includegraphics[width = 3.3in]{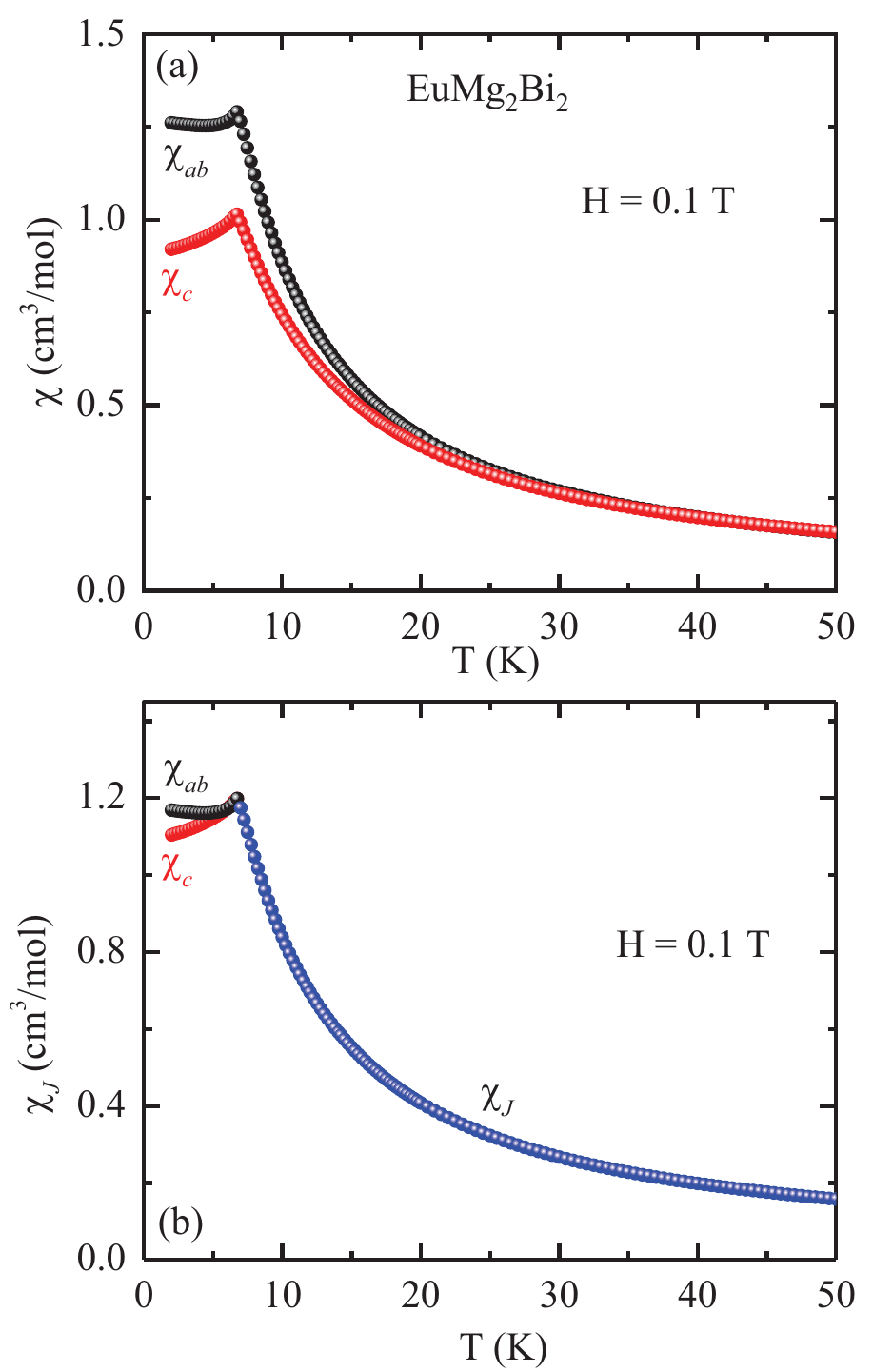}
\caption{(a) Temperature~$T$ dependence of the zero-field-cooled (ZFC) magnetic susceptibility~$\chi$ of \emb\ measured in a magnetic field $H = 0.1$~T applied in the $ab$~plane ($\chi_{ab}$) and along the $c$~axis $(\chi_c)$. (b) The Heisenberg susceptibility $\chi_J$ in the PM state ($T > T_{\rm N}$) shown as blue symbols was calculated by taking the spherical average of the data in~(a) at $T\geq T_{\rm N}$ using Eq.~(\ref{Eq.Heisenbergsusceptibility}). The $\chi_{ab}$ and $\chi_c$ data at $T \leq T_{\rm N}$ in~(a) were then vertically adjusted in~(b) to match the $\chi_J(T)$ data at $T_{\rm N}$.}
\label{Chi_T_1kOe}
\end{figure}

The magnetic susceptibility $\chi = M/H$ of a \emb\ single crystal was measured at different applied magnetic fields~$H$ in both zero-field-cooled (ZFC) and field-cooled (FC) modes. However, no signature of thermal hysteresis could be evidenced down to the lowest measured temperature even for the lowest applied magnetic field of 0.05~T\@. The temperature dependences of $\chi$ in ZFC mode measured at $H=0.1$~T applied in the $ab$~plane ($H \parallel ab$) and along the $c$~axis ($H \parallel c$) are shown in Fig.~\ref{Chi_T_1kOe}(a). Both $\chi_{ab}$ and $\chi_{c}$ exhibit a sharp peak at $T_{\rm N} = 6.70(5)$~K for $H = 0.1$~T which is close to the previously reported $T_{\rm N}$ for the compound~\cite{May2011, Ramirez2015, Kabir2019}. Below $T_{\rm N}$, both $\chi_{ab}$ and $\chi_c$ are almost independent of $T$\@.  Because $\chi_{c}<\chi_{ab}$ below $T_{\rm N}$ as in Fig.~\ref{Chi_T_1kOe}(a), it was suggested previously that the Eu ordered moments with spins-7/2 aligned antiferromagnetically along the $c$~axis. We obtain a different model below. 

$\chi_{ab}$ and $\chi_{c}$ start to diverge from each other below \mbox{$T \approx 40$~K}, which is much higher than $T_{\rm N}$.  This divergence suggests the occurrence of anisotropic FM fluctuations below this temperature, with the strongest fluctuations in the $ab$~plane. We removed the influence of magnetic anisotropy in the paramagnetic (PM) state above $T_{\rm N}$ by carrying out a spherical average of the data according to
\bea
\chi_{J}(T)=\frac{1}{3}[2\chi_{ab}(T)+\chi_{c}(T)] \qquad (T\geq T_{\rm N})
\label{Eq.Heisenbergsusceptibility}
\eea
as shown in Fig.~\ref{Chi_T_1kOe}(b), where the designation $\chi_J$ denotes that the remaining deviation of the temperature dependence of $\chi$ from Curie-Weiss behavior at $T\geq T_{\rm N}$  arises from exchange interactions~$J$\@.  Then the $\chi_{ab}(T)$ and $\chi_c(T)$ data at $T\leq T_{\rm N}$ were respectively shifted vertically so that these susceptibilities matched the spherically-averaged PM susceptibility at $T_{\rm N}$ as shown in Fig.~\ref{Chi_T_1kOe}(b).

Interestingly, Fig.~\ref{Chi_T_1kOe}(b) shows that after correcting for the anisotropy in $\chi$ above $T_{\rm N}$, $\chi_{ab}$ and $\chi_{c}$ are nearly the same and nearly independent of $T$ below $T_{\rm N}$\@. According to MFT~\cite{Johnston2012, Johnston2015}, these two features suggest that the Eu spins are ordered in either a stacked planar 120$^{\circ}$ configuration or in a $c$-axis helix with turn angle $kd \approx 120^\circ$. Such behavior was previously observed for different $120^\circ$-ordered triangular-lattice AFM systems~\cite{Katsufuji2001, Brown2006, Maruyama2001} including for the most quantum spin $S=1/2$, but where the explanation was not available at that time.  Indeed, the MFT predicts that for 120$^\circ$ ordering in either a planar structure or in a helix with a $120^\circ$ turn angle between layers, below $T_{\rm N}$ the susceptibility should be isotropic and independent of both $T$ and the value of the spin quantum number~$S$~\cite{Johnston2012, Johnston2015}.

\begin{figure}
\includegraphics[width = 3.3in]{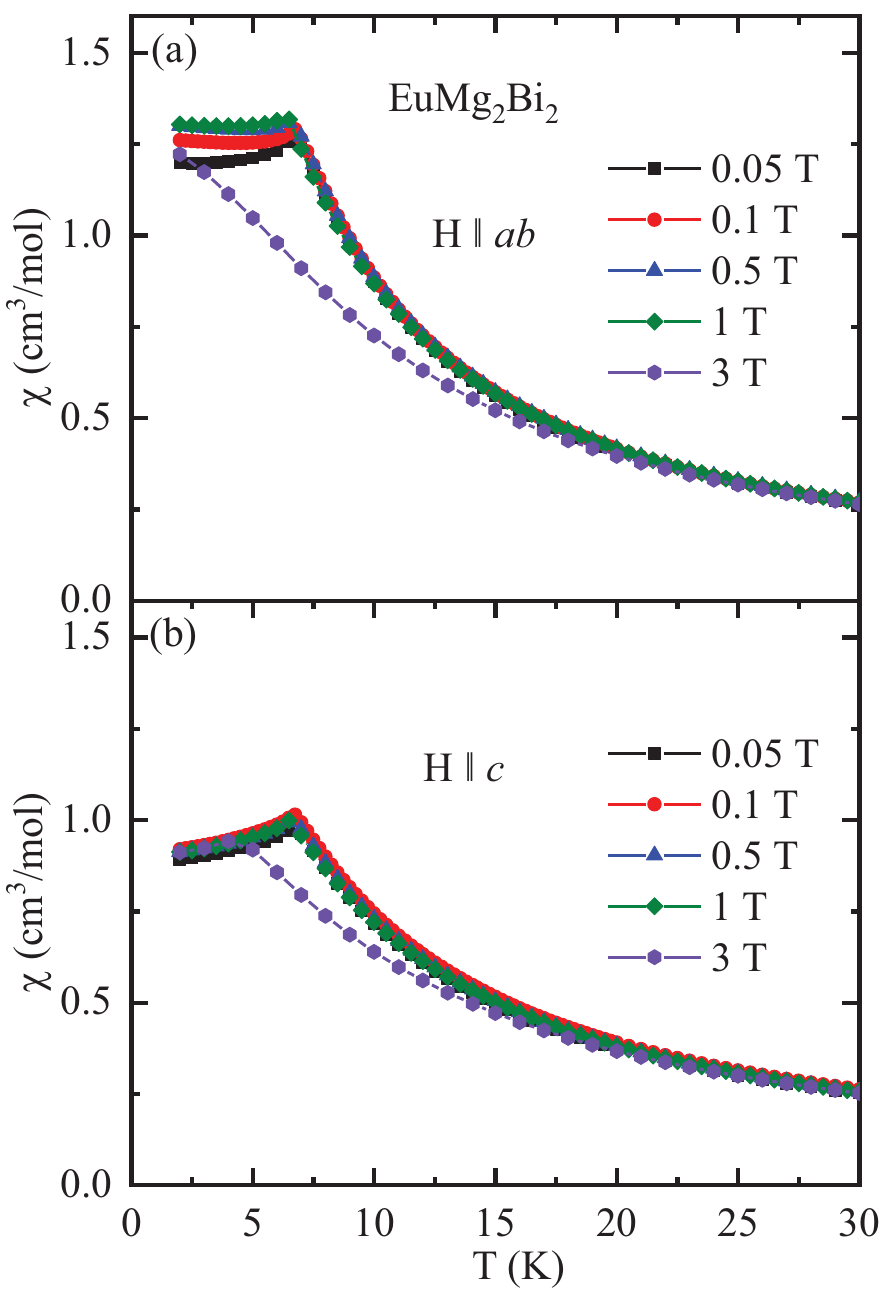}
\caption{Zero-field-cooled magnetic susceptibility versus temperature~$T$ measured at different applied magnetic fields $H$ as listed for (a)~$H\parallel ab$ and (b)~$H\parallel c$.}
\label{Chi_T_Diff_Field}
\end{figure}

The anisotropic field-dependent magnetic susceptibilities $\chi = M/H$ for $H\parallel ab$ and $H\parallel c$ are plotted versus~$T$ for $T<30$~K in Figs.~\ref{Chi_T_Diff_Field}(a) and~(b), respectively. The $T_{\rm N}$ shifts to lower temperature with increasing $H$ for both field directions, but the suppression for $H>1$~T is clearly faster for $H\parallel ab$ than for $H\parallel c$.

\subsection{Inverse magnetic susceptibility and Curie-Weiss behavior in the paramagnetic state}

\begin{figure}
\includegraphics[width = 3.3in]{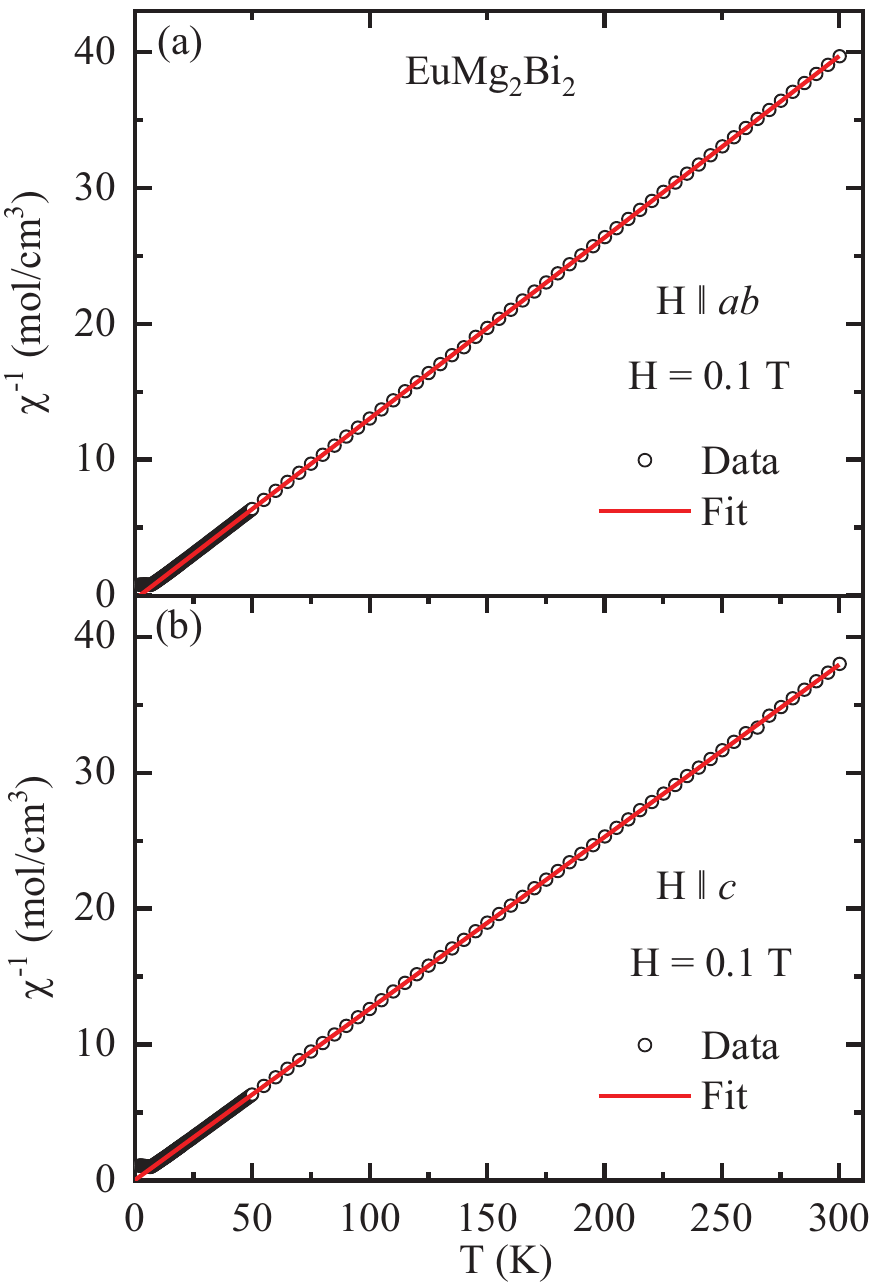}
\caption{Inverse magnetic susceptibility $\chi^{-1}(T)$ versus temperature $T$ for $H = 0.1$~T for (a) $H\parallel ab$ and (b)~$H\parallel c$ along with the respective fits by the modified Curie-Weiss law~(\ref{Eq.ModCurieWeiss}).}
\label{Chi-1_T_1kOe}
\end{figure}
The inverse magnetic susceptibility data in the PM region are fitted by the modified Curie-Weiss law
\bea
\chi_{\alpha}(T) =\chi_0 + \frac{C_{\alpha}}{T-\theta_{\rm p\alpha}} \qquad (\alpha ~=~ab,~c),
\label{Eq.ModCurieWeiss}
\eea
where $\chi_0$ is an isotropic temperature-independent term, $\theta_{\rm p}$ is the Weiss temperature, and $C$ is the Curie constant given by
\bea
C_{\alpha}&=&\frac{N_{\rm A} {g_\alpha}^2S(S+1)\mu^2_{\rm B}}{3k_{\rm B}} = \frac{N_{\rm A}\mu^2_{\rm {eff, \alpha}}}{3k_{\rm B}},
\label{Eq.Cvalue1}
\eea
where $N_{\rm A}$ is Avogadro's number, $g_\alpha$ is the spectroscopic splitting factor ($g$ factor) for the $\alpha^{\rm th}$ direction, $k_{\rm B}$ is Boltzmann's constant, and $\mu_{\rm eff}$ is the effective moment of an Eu spin in units of Bohr magnetons $\mu\rm_B$. Figures~\ref{Chi-1_T_1kOe}(a) and \ref{Chi-1_T_1kOe}(b) depict the $\chi^{-1}(T)$ behavior in $H$ = 0.1 T for $H \parallel ab$ and $H \parallel c$, respectively, along with the modified Curie-Weiss fits obtained using Eq.~(\ref{Eq.ModCurieWeiss}). The parameters of the fits for the two different field directions are listed in Table~\ref{Tab.chidata}. The $\mu_{\rm eff}$ values obtained from $C$ for both applied field directions are close to the value~$7.94\,\mu_{\rm B}$ expected for Eu$^{+2}$ spins $S = 7/2$ with $g$ = 2. The positive value of $\theta_{ab}$ indicates predominant FM in-plane  interactions. The difference in magnitude of $\theta_{ab}$ and $\theta_{c}$ may be due to the increased magnitude of FM fluctuations in the $ab$ plane compared to those along the $c$~axis as discussed above.

\begin{table}
\caption{\label{Tab.chidata} The $T$--independent contribution to the susceptibility $\chi_0$, Curie constant per mol $C_\alpha$ for fields in the $\alpha = ab, c$ directions, effective moment per Eu $\mu{\rm_{eff}(\mu_B/Eu)} = \sqrt{8C}$ and Weiss temperature $\theta\rm_{p\alpha}$ obtained from the $\chi^{-1}(T)$ versus $T$ fits for $H = 0.1$~T using Eq.~(\ref{Eq.ModCurieWeiss}).}
\begin{ruledtabular}
\begin{tabular}{ccccc}	
Field & $\chi_0$ 				& $C_{\alpha}$ 		    &  $\mu_{\rm eff\alpha}$ 	& $\theta_{\rm p\alpha}$ \\
direction	 	& $\rm{\left(10^{-4}~\frac{cm^3}{mol}\right)}$	 & $\rm{\left(\frac{cm^3 K}{mol}\right)}$    & $\rm{\left(\frac{\mu_B}{Eu}\right)}$& (K)  \\
\hline
$H\parallel ab$ 		& $-1.2(3)$		&  	7.538(9)	&	7.766(4)		& 2.07(9)		\\
$H\parallel c$ 		& $-0.2(2)$ 		&  	7.948(22)	&	7.97(1)		& $-0.2(2)$	\\
\end{tabular}
\end{ruledtabular}
\end{table}

\subsection{Heisenberg exchange interactions from MFT model}

The Heisenberg exchange interactions between the Eu spins were estimated using a minimal one-dimensional $J_0$-$J_1$-$J_2$ MFT model for a helix~\cite{Johnston2019_J0J1J2,Nagamiya_1967} (see Fig.~\ref{Fig:Fig_Helix}).  Here $J_0$ is the sum of the Heisenberg exchange interactions of a representative spin with all other spins in the same $ab$-plane layer,  $J_1$ is the sum of exchange interactions of a spin with all spins in a nearest layer along the helix axis, and $J_2$ is the sum of the exchange interactions of a spin with all spins in a next-nearest layer (see Fig.~\ref{Fig:Fig_Helix}).  According to this MFT model, these exchange interactions are related to the turn angle $kd$, AFM transition temperature $T_{\rm N}$, and Weiss temperature~$\theta_{\rm p}$ by~\cite{Johnston2012, Johnston2015}
\bse
\label{Eqs:J0J1J2}
\bea
&&\cos(kd) = -\frac{J_{1}}{4J_{2}},\\*
T_{\rm N} &=& -\frac{S(S+1)}{3k_{\rm B}} \big[J_0 + 2J_{1}\cos(kd)\nonumber\\*
&& \hspace{0.9in} +\ 2J_{2}\cos(2kd)\big], \label{eq:TN}\\*
\theta_{\rm p} &=& -\frac{S(S+1)}{3k_{\rm B}} \left(J_0+2J_{1}+2J_{2}\right),
\label{eq:thetap}
\eea
\ese
where a positive (negative) $J$ corresponds to net AFM (FM) interactions.  The values of $J_0$, $J_1$, and $J_2$ were estimated using the parameters $S = 7/2,\ T_{\rm N} = 6.7\ {\rm K},\ \theta_{\rm p}=\theta_{\rm p\,ave}= 1.31(13)$~K (spherical average), and $kd=0.66\pi\approx 120^\circ$, yielding
\bse
\label{Eqs:J0J1J2values}
\bea
J_0/k_{\rm B} = -0.934(8)~\rm{K (FM)},\\*
J_1/k_{\rm B} = 0.228(6)~\rm{K (AFM)},\\*
J_2/k_{\rm B} = 0.114(2)~\rm{K (AFM)}.
\label{Eq.J012values}
\eea
\ese
The FM value of $J_0$ and AFM values of $J_1$ and $J_2$ are consistent with a $c$-axis helical spin structure with the moments ferromagnetically-aligned in each $ab$~plane layer~\cite{Johnston2012, Johnston2015}.  This model is also consistent with the Weiss-temperature anisotropy in Table~\ref{Tab.chidata}.  The alternative possibility of a stacked triangular-lattice planar array of ordered moments with an angle of $\approx 120^\circ$ between adjacent moments in each layer is less likely because in that case one would expect $\theta_{ab}$ to be AFM (negative) instead of positive (FM) as given in Table~\ref{Tab.chidata}.

\subsection{Magnetization versus applied magnetic field isotherms}

\begin{figure}
\includegraphics[width = 3.3in]{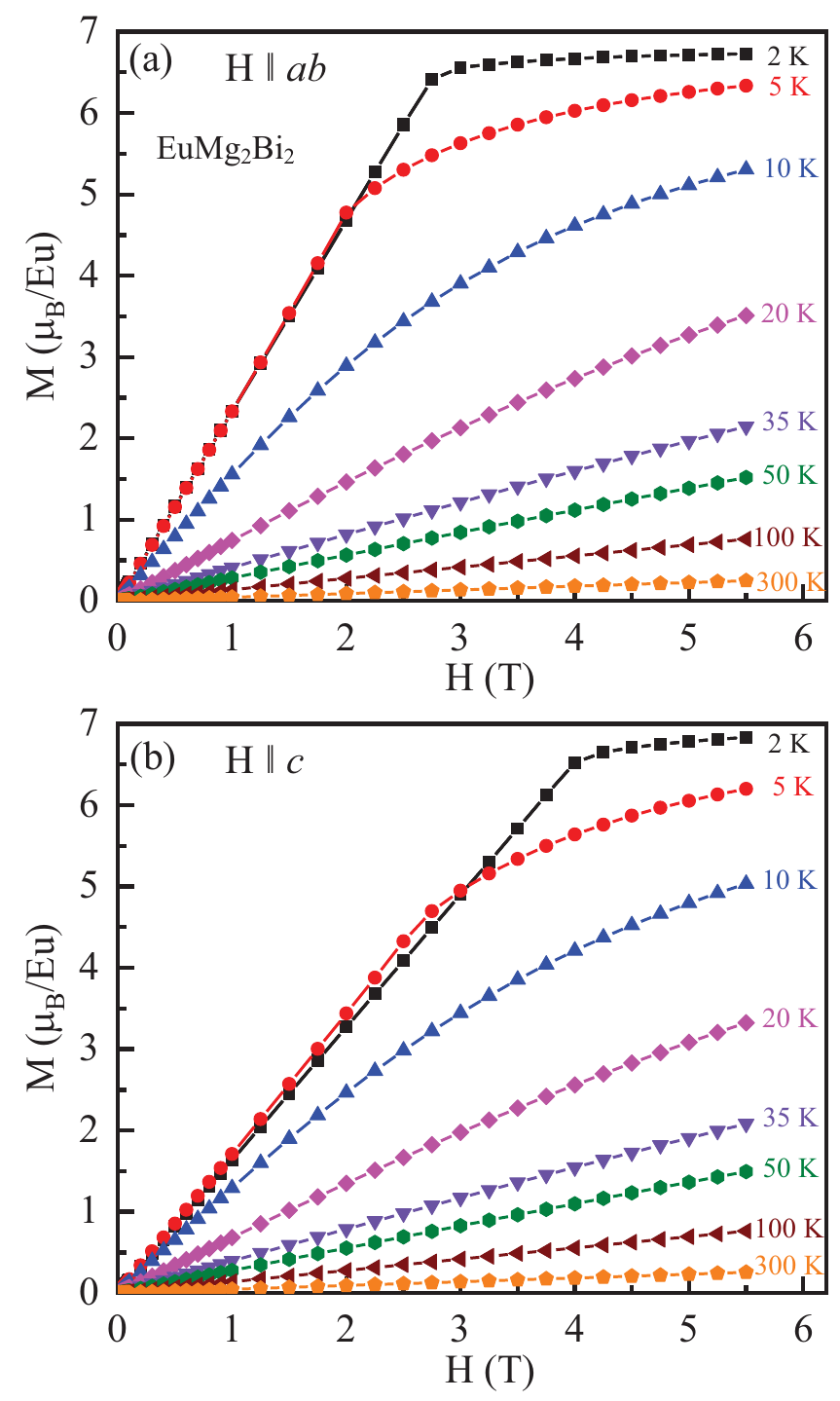}
\caption{Field dependence of magnetic isotherms measured at different temperatures when the applied field is (a) in the $ab$ plane and (b) along the $c$~axis.}
\label{MH_All}
\end{figure}

\begin{figure}
\includegraphics[width = 3.3in]{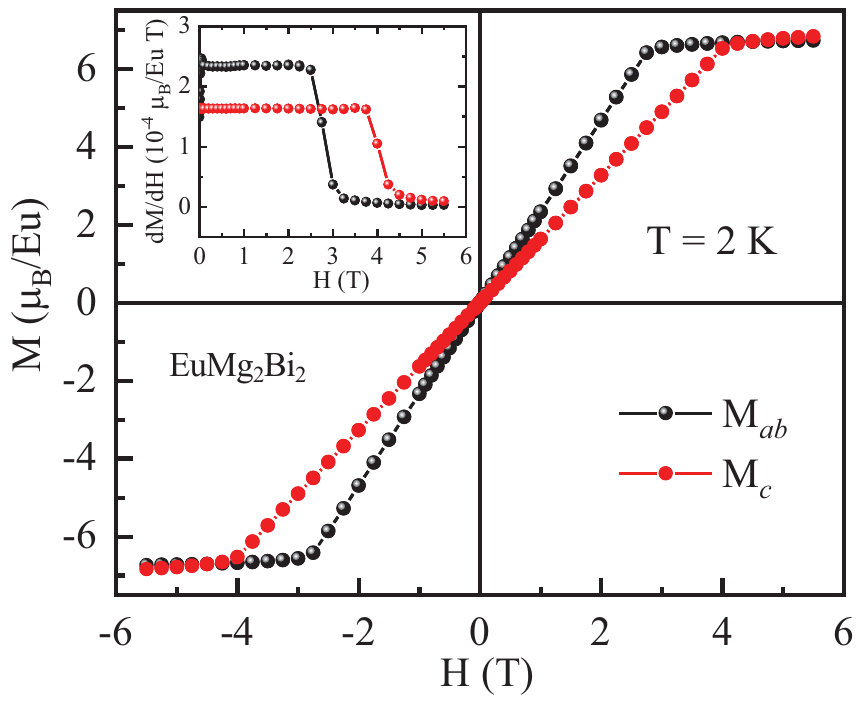}
\caption{Magnetization $M_{ab}$ and $M_{c}$ as a function of the applied field $H$ at temperature $T = 2$~K\@. Inset: $dM/dH$ versus $H$ at $T$ = 2 K for both field directions illustrating in more detail the behavior near the respective critical fields.}
\label{MH_2K}
\end{figure}

\begin{figure}
\includegraphics[width = 3.in]{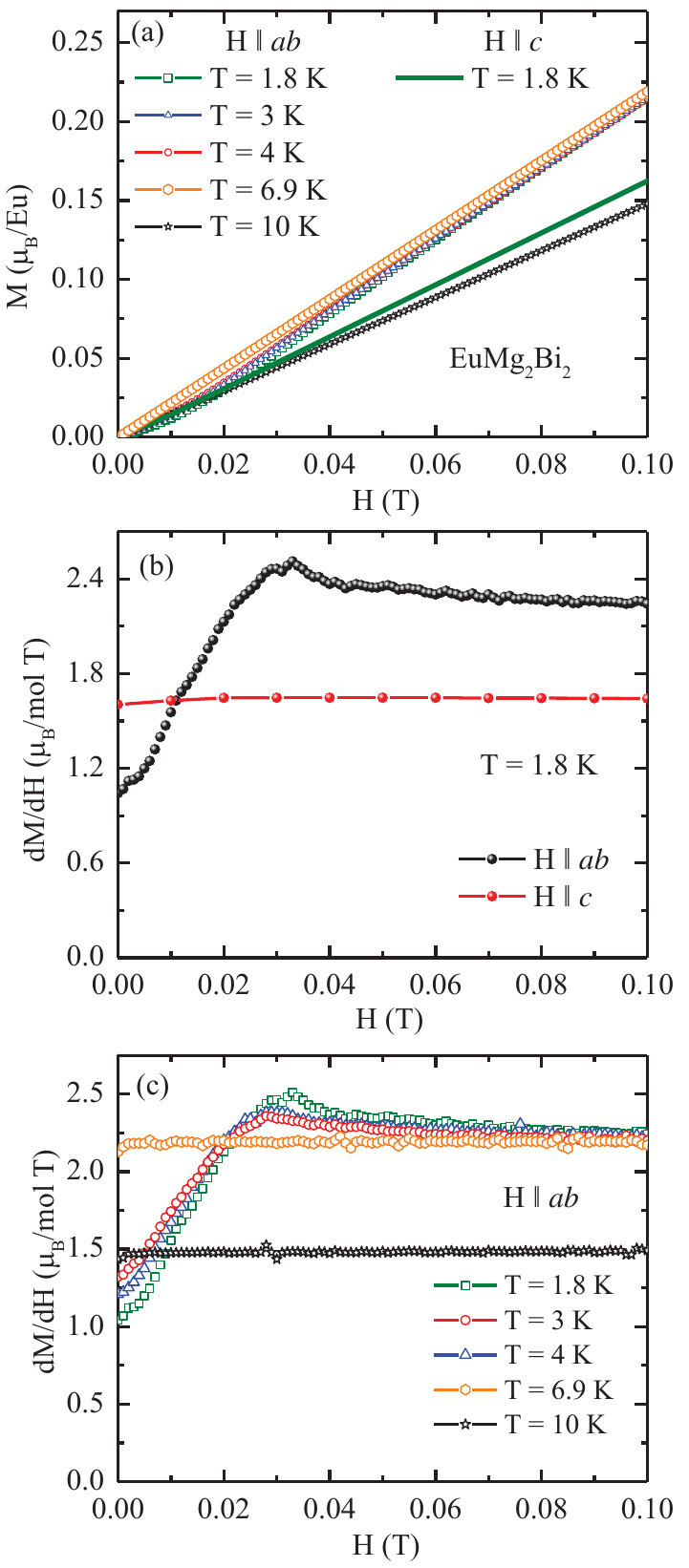}
\caption{(a) Magnetization versus field $M(H)$ isotherms at low fields for different temperatures with $H \parallel ab$ and $H \parallel c$. (b)~The derivative $dM/dH$ versus $H$ at $T = 1.8$~K for both field directions. (c)~$dM/dH$ versus~$H$ at different temperatures for $H \parallel ab$.}
\label{Fig:dMdH_low_field}
\end{figure}

Isothermal magnetization $M$ versus~$H$ data measured at different temperatures for $H \parallel ab$ and $H \parallel c$ are shown in Figs.~\ref{MH_All}(a) and~\ref{MH_All}(b), respectively. The $M(H)$ data at $T = 2$~K are shown in Fig.~\ref{MH_2K} separately for the two different field directions for comparison. On the scale of the figures, $M_{ab}(H)$ and $M_{c}(H)$ at 2~K increase linearly with increasing $H$ up to the respective critical fields $H_{{\rm c}\,ab} = 2.75(2)$~T and $H_{{\rm c}\,c} = 4.0(3)$~T, above which the crystal enters the PM state and $M$ begins to saturate. At $T=2$~K, $M_{ab}(H)$ and $M_{c}(H)$ attain values of $M_{\rm c} \approx 6.80(5)~\mu_{\rm B}$/Eu at $H=5.5$~T\@.  In view of the data in Table~\ref{Tab.chidata} obtained by fitting the $\chi(T)$ data, the $M(H)$ data for both field directions should indeed saturate to about $7~\mu_{\rm B}$/Eu at sufficiently high fields.  Hysteresis in $M(H)$ was not observed for either of the two applied field directions (not shown), consistent with the magnetic structure not having a glassy or FM component.  These results are also consistent with the predictions of MFT for a $c$-axis helix with a turn angle of 120$^\circ$~\cite{Johnston2015, Johnston2017, Johnston2017_2}.  For the in-plane $M_{ab}(H)$ with this turn angle, a smooth crossover between a helix and a fan phase is predicted even though the theoretical prediction in this case is that $M\propto H$ below saturation.  This prediction is indeed in apparent agreement with the data in Fig.~\ref{MH_All}(a) and the low-field data at $T=1.8$~K displayed in Fig.~\ref{Fig:dMdH_low_field}(a).

However, a detailed analysis of these $M_{ab}(H)$ data reveal a nonlinearity with a peak at $H \approx 300$~Oe from the derivative plot $dM_{ab}/dH$ versus~$H$ in Fig.~\ref{Fig:dMdH_low_field}(b) that is not predicted by the MFT, whereas the corresponding $c$-axis data in this figure show no such feature in agreement with MFT\@.  The deviation from linearity in the low-$H$ $M_{ab}(H)$ data is also temperature dependent as seen from Fig.~\ref{Fig:dMdH_low_field}(c). The deviation is strongest at $T$ = 1.8 K, diminishes with increasing~$T$, and vanishes for $T > T_{\rm N}$.  All these isothermal low-$H$ $M(H)$ measurements were carried out on warming after quenching the superconducting magnet to avoid remanent fields in the magnet that might affect the results.  Although not predicted by MFT for $kd=2\pi/3$~rad, the nonlinearity of $M_{ab}(H)$ is consistent with expectation for a noncollinear AFM structure with the moments aligned in the $ab$~plane.  It would be interesting to investigate by neutron diffraction how the magnetic structure changes in the low-field region below 1000~Oe.


\section{\label{Sec:Cp} Heat capacity and $H_\perp$-$T$ Phase Diagram}

\begin{figure}
\includegraphics[width = 3.3in]{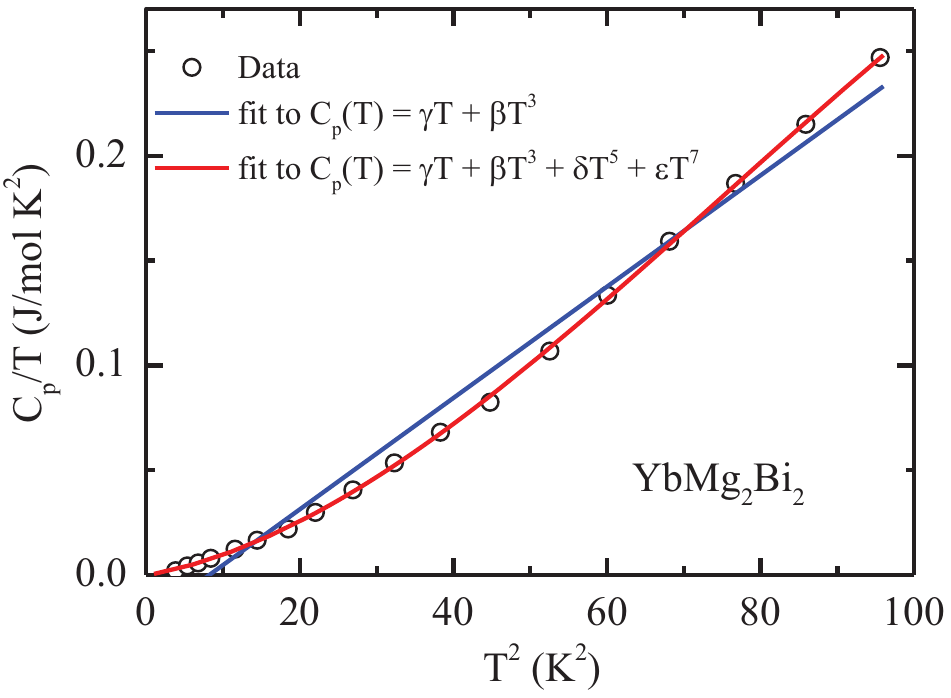}
\caption{Low-temperature heat capacity $C_{\rm p}/T$ versus temperature $T$ of \ymb\ from 2 to 10~K\@.  Two fits are shown. The two-term fit is the conventional one in Eq.~(\ref{Eq:2TermFit}) containing the Sommerfeld coefficient $\gamma$ and a single Debye lattice heat capacity term.  This behavior obviously does not fit the data.  The four-term fit in Eq.~(\ref{Eq:4TermFit}) with the fitted parameters in Eq.~(\ref{Eq:4TermFitPars}) includes two additional lattice heat-capacity terms.  }
\label{Cp_T_YbMg2Bi2}
\end{figure}

In this section we use the heat capacity $C_{\rm p}(T)$ of the nonmagnetic compound \ymb\ as a reference for the lattice heat capacity of \emb.  The data for single-crystal \ymb\ at low~$T$ are shown in Fig.~\ref{Cp_T_YbMg2Bi2}, plotted as $C_{\rm p}/T$ versus~$T^2$\@.  For most nonmagnetic materials, the behavior at low temperatures is described by the expression
\bea
\frac{C_{\rm p}}{T} = \gamma + \beta T^2,
\label{Eq:2TermFit}
\eea
where $\gamma$ is the Sommerfeld coefficient associated with degenerate itinerant charge carriers and $\beta$ is the coefficient of the $T^3$ term in the low-$T$ limit of the Debye lattice heat capacity.  However, from Fig.~\ref{Cp_T_YbMg2Bi2} this two-term fit does not fit the data at all.  Furthermore, it yields an unphysical negative value for $\gamma$.  In order to obtain a good fit to the data below 10~K we added two additional lattice heat capacity terms according to
\bea
\frac{C_{\rm p}}{T} = \gamma + \beta T^2 + \delta T^4 + \varepsilon T^6.
\label{Eq:4TermFit}
\eea
An excellent fit to the data was obtained by this expression as illustrated in Fig.~\ref{Cp_T_YbMg2Bi2} where
\bea
\gamma = -0.2(9)~{\rm \frac{mJ}{mol\,K^2}},\quad \beta = 0.6(1)~{\rm \frac{mJ}{mol\,K^4}}, \label{Eq:4TermFitPars}\\*
\delta = 35(3)~{\rm \frac{\mu J}{mol\,K^6}},\quad \varepsilon = -0.15(2)~{\rm \frac{\mu J}{mol\,K^8}}.\nonumber
\eea
The value of $\Theta_{\rm D}$ is obtained from $\beta$ according to
\bea
\Theta_{\rm D} = \left(\frac{12\pi^4nR}{5\beta}\right)^{1/3},
\label{QDcalc}
\eea
where $n$ is the number of atoms per formula unit ($n=5$ here) and $R$ is the molar gas constant, yielding $\Theta_{\rm D} = 255(15)$~K\@.  It is notable that $\gamma=0$ to within its error.  Therefore below we assume $\gamma=0$ when evaluating the heat capacity of \emb.

\begin{figure}
\includegraphics[width = 3in]{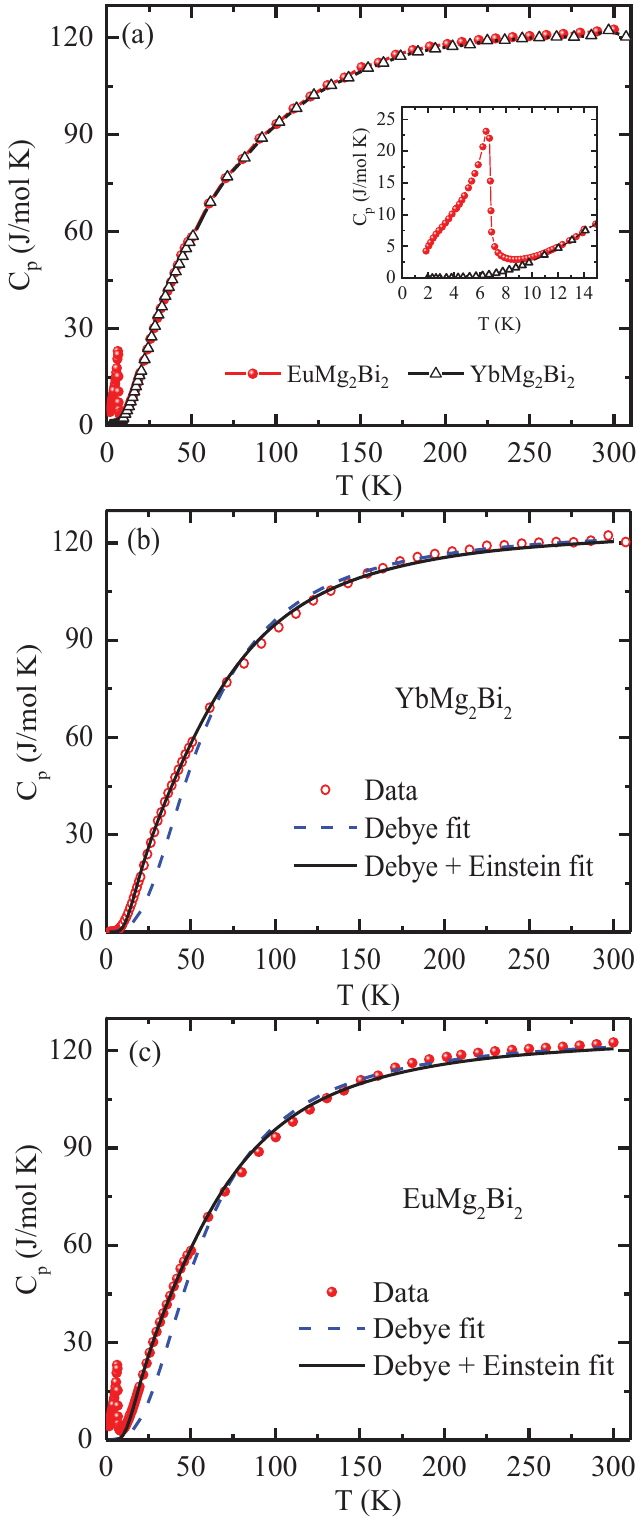}
\caption{(a) Zero-field heat-capacity $C_{\rm p}$ versus temperature~$T$ of \emb\ and \ymb\ single crystals. The data for the two compounds are hardly distinguishable above 20~K\@. Inset:~expanded low-temperature region around $T_{\rm N}$ of \emb. The lines are guides to the eye.  $C_{\rm p}(T)$ of (b)~\ymb\ and (c)~\emb\ along with the respective fits by the Debye model (dashed curves) and by the sum of Debye and Einstein terms (solid curves).}
\label{Cp_T_zero_field}
\end{figure}

\begin{table}
\caption{\label{Tab:LattCpFitdata}  Fitting parameters in Eqs.~(\ref{Eq:Debye_Fit})--(\ref{Eq:Einstein}) from fits of the lattice heat capacities of \ymb\ from 2 to 300~K and \emb\ from 25 to 300~K\@.}
\begin{ruledtabular}
\begin{tabular}{cccc}	
Compound  			& $\Theta_{\rm D}$		&  $\Theta_{\rm E}$ 	& $\alpha$	 \\
		 			& (K)	 				& (K)    			&  			\\
\hline
\ymb\ 				& 309(5)				&  75(2)			& 0.37(1)		\\
\emb\ 				& 305(10)				&  77(5)			& 0.38(3)	 	\\
\end{tabular}
\end{ruledtabular}
\end{table}

The zero-field single-crystal heat capacities $C_{\rm p}(T)$ of \emb\ and the nonmagnetic analogue \ymb\ measured in the temperature range 1.8--300~K are plotted in Fig.~\ref{Cp_T_zero_field}(a). The $C_{\rm p}$ of \emb\ exhibits a pronounced peak at $T_{\rm N}$ = 6.7 K as evident from the inset of Fig.~\ref{Cp_T_zero_field}(a).  The $C_{\rm p}(T)$ attains values of \mbox{$\approx 122.5$~J/mol~K} and $\approx 121$~J/mol K at $T = 300$~K for \emb\ and \ymb, respectively. These values are close to the classical Dulong-Petit limit due to acoustic phonon modes $C_{\rm V} = 3nR = 124.7$~J/mol~K with $n=5$ being the number of atoms per formula unit of the above compounds.

We fitted the $C_{\rm p}(T)$ data in the temperature regions 25--300~K for \emb\ and 2--300~K for \ymb\ by the Debye lattice heat capacity prediction
\bea
C_{\rm p}(T) &=& nC_{\rm V\,Debye}(T),\label{Eq:Debye_Fit} \\*
C_{\rm V\,Debye}(T) &=& 9R \left(\frac{T}{\Theta_{\rm D}}\right)^3\int_{0}^{\Theta_{\rm D}/T}\frac{x^4e^x}{(e^x-1)^2} dx,\nonumber
\eea
where $\Theta_{\rm D}$ is the Debye temperature. The Pad\'e approximant for the Debye function in Ref.~\cite{Goetsch_2012} was used for the fits.  As seen from dashed curves in Figs.~\ref{Cp_T_zero_field}(b) and \ref{Cp_T_zero_field}(c), the lattice heat capacity is not described well by the Debye model. Much better fits were obtained by including an Einstein lattice contribution to the fits according to
\bea
\label{Eq:Debye and Einstein}
C_{\rm p}(T) &=& (1 - \alpha)C_{\rm V\,Debye} + \alpha C_{\rm V\,Einstein},
\eea
where
\bea
\label{Eq:Einstein}
C_{\rm V\,Einstein}(T) &=& 3R \left(\frac{\Theta_{\rm E}}{T}\right)^2\frac{e^{\Theta_{\rm E}/T}}{(e^{\Theta_{\rm E}/T} - 1)^2}
\eea
with $\Theta_{\rm E}$ being the Einstein temperature. The parameter $\alpha$ determines the fraction of the Einstein  contribution to the total lattice heat capacity. Very good fits of the $C_{\rm p}(T)$ data by Eq.~(\ref{Eq:Debye and Einstein}) were achieved, as depicted by the solid curves in Figs.~\ref{Cp_T_zero_field}(b) and~\ref{Cp_T_zero_field}(c). The fitted parameters are listed in Table~\ref{Tab:LattCpFitdata}. The Debye temperatures of $\sim 300$~K are much larger than the values of 207~K and 211~K previously reported from Debye fits to the data from 20 to 200~K for \ymb\ and \emb, respectively~\cite{May2011}.  The low Einstein temperatures of $\sim 75$~K suggest the presence of low-frequency optic modes associated with the heavy Bi and/or Yb or Eu atoms, respectively, as also suggested from Fig.~\ref{Cp_T_YbMg2Bi2}.

\begin{figure}
\includegraphics[width = 3.in]{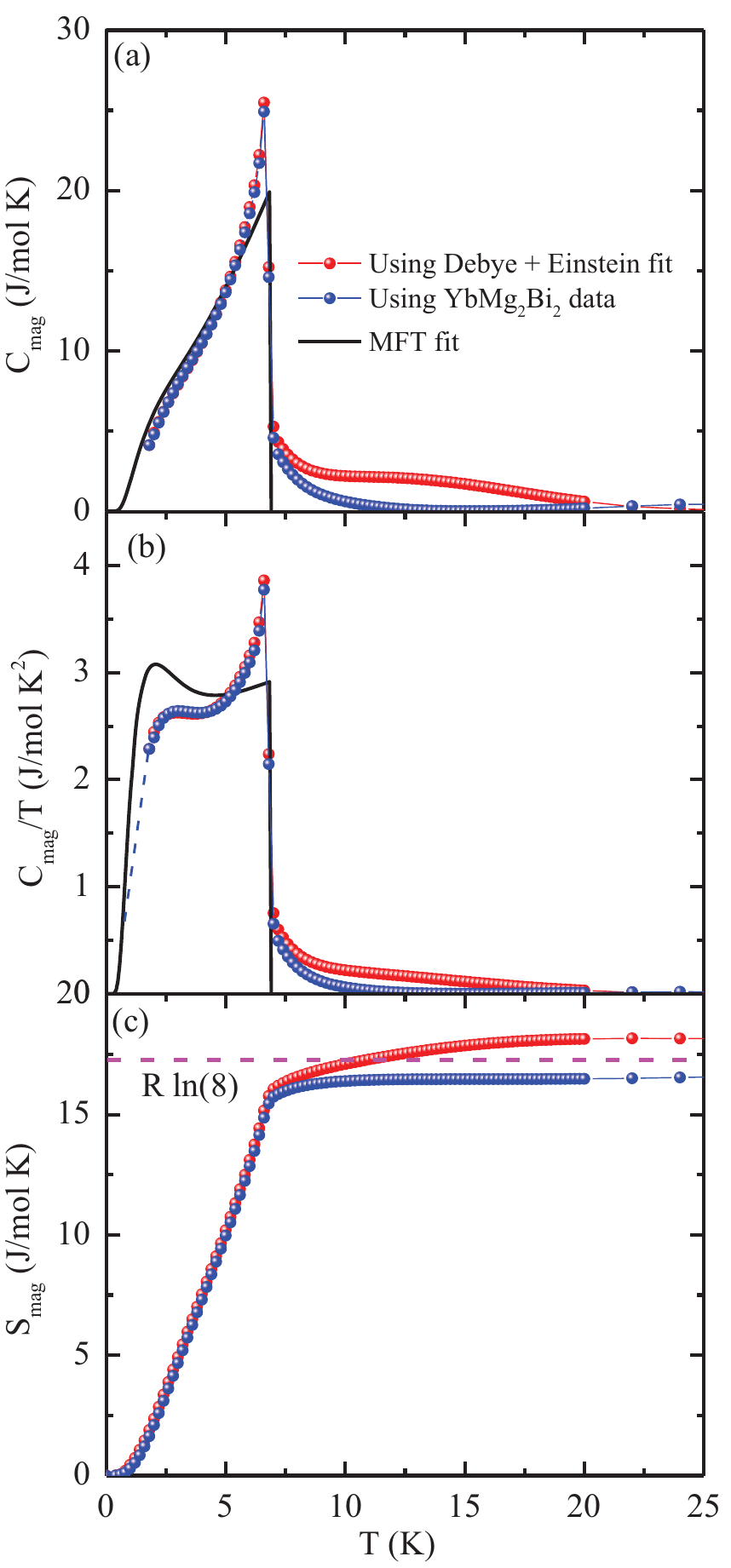}
\caption{Temperature $T$ dependence of (a)~the magnetic component of the heat capacity $C_{\rm {mag}}$, (b)~$C_{\rm {mag}}/T$, and (c)~the magnetic entropy $S_{\rm {mag}}$ obtained from the $C_{\rm mag}(T)/T$ data in~(b) using Eq.~(\ref{Eq:Entropy}).  The solid black lines in (a) and~(b) represent the predictions of MFT in Eqs.~(\ref{Eqs:Cmag}) for $S=7/2$ and $T_{\rm N} = 6.7$~K\@.  In (b), the experimental data for $T\geq 2$~K were extrapolated to $T=0.7$~K using a spline fit  where it intersected the MFT prediction as shown by the dashed blue line and then the MFT prediction was used below 0.7~K\@.  }
\label{Cmag_T_zero_field}
\end{figure}

The magnetic contribution $C_{\rm {mag}}(T)$ to the heat capacity of \emb\ was obtained by subtracting the $C_{\rm p}(T)$ data of the nonmagnetic analogue \ymb\ in Fig.~\ref{Cp_T_zero_field}(a) after correcting for the difference in formula weights, where the temperature scale~$T^*$ of \ymb\ was obtained as
\bea
\label{Eq:FWCorrection}
T^{\ast} = \frac{T}{(M_{\rm EuMg_2Bi_2}/M_{\rm YbMg_2Bi_2})^{1/2}},
\eea
where $T$ is the experimentally-measured temperature. The $C_{\rm {mag}}(T)$ was also estimated by subtracting the lattice contribution obtained from the fit of the $C_{\rm p}(T)$ data by Eq.~(\ref{Eq:Debye and Einstein}). Figure~\ref{Cmag_T_zero_field}(a) shows $C_{\rm {mag}}(T)$ obtained using the two different lattice heat capacity estimates. The two calculations of $C_{\rm mag}(T)$ agree very well below $T_{\rm N}$, but the data above $T_{\rm N}$ obtained using the measured lattice contribution of \ymb\ is physically more likely.
 
According to MFT, the discontinuity in $C_{\rm {mag}}$ at \mbox{$T = T_{\rm N}$} is given by~\cite{Johnston2015}
\bea
\label{Eq:deltaCp}
\Delta C{\rm_{mag}}=\frac{5S(1+S)}{1+2S+2S^2}R.
\label{Eq:DeltaCmag}
\eea
The jump $\Delta C_{\rm mag}$ in $C_{\rm {mag}}$ for \emb\ at $T_{\rm N}$ = 6.7 K expected from Eq.~(\ref{Eq:DeltaCmag}) is 20.14~J/mol~K for $S = 7/2$. The experimentally-observed jump $\Delta C_{\rm {mag}}(T_{\rm N})\approx 25$~J/mol~K is significantly larger than the MFT prediction. This difference arises because the observed $\lambda$ shape of $C_{\rm mag}(T)$ below $T_{\rm N}$ is different from step shape of the MFT prediction shown as the solid curve in Fig.~\ref{Cmag_T_zero_field}(a) obtained as follows.


The magnetic contribution $C_{\rm mag}(T,H=0)$ to the heat capacity according to MFT is given by~\cite{Johnston2015}
\bse
\label{Eqs:Cmag}
\bea
\label{Eq:deltaCp_T_MFT}
\frac{C_{\rm {mag}}(t)}{R} &=& \frac{3S\bar{\mu}_0^2(t)}{(S + 1)t[\frac{(S + 1)t}{3B^{\prime}_S[y_0(t)]} - 1]},\\
\bar{\mu}_0(t) &=& B_S[y_0(t)],\\
y_0(t) &=& \frac{3\bar{\mu}_0(t)}{(S+1)t},\\
t &=& T/T_{\rm N}(H=0),
\eea
\ese
where $\bar{\mu}_0(t)\equiv \mu_0(t)/\mu_{\rm sat}$ is the reduced ordered moment versus $t$ in $H = 0$, $\mu_{\rm sat} = gS\mu_{\rm B}$ is the saturation moment of the spin, $B_S(y)$ is Brillouin function, and $B^{\prime}_S(y_0) \equiv \frac{dB_S(y)}{dy}\big|_{y=y_0}$. The solid black curve in Fig.~\ref{Cmag_T_zero_field}(a) represents the $C_{\rm {mag}}(T)$ predicted by MFT for $T_{\rm N} = 6.7$~K and $S = 7/2$.  The disagreements with the data in Fig.~\ref{Cmag_T_zero_field}(a) arise from neglect of dynamic magnetic fluctuations and correlations in MFT\@.

The zero-field magnetic entropy $S_{\rm {mag}}(T)$ for \emb\ in Fig.~\ref{Cmag_T_zero_field}(c) was obtained from the zero-field $C_{\rm {mag}}(T)/T$ data in Fig.~\ref{Cmag_T_zero_field}(b) using the relation
\bea
\label{Eq:Entropy}
S_{\rm mag}(T) = \int_{0}^{T}\frac{C_{\rm {mag}}(T^\prime)}{T^\prime} dT^\prime.
\eea
The $C_{\rm {mag}}(T)$ data in the $T$ range 0--1.8 K for which we have no data was estimated as described in the caption to Fig.~\ref{Cmag_T_zero_field} and shown as the dashed blue line in Fig.~\ref{Cmag_T_zero_field}(b). The $S_{\rm {mag}}(T)$ calculations for \emb\ using the above two methods of calculating $C_{\rm {mag}}(T)$ are shown in Fig.~\ref{Cmag_T_zero_field}(c). The $S_{\rm {mag}}(T)$ calculated using the Debye-Einstein lattice contribution saturates at high~$T$ to a value of 18.2 J/mol~K, which is slightly larger than the theoretical high-$T$ limit $S_{\rm mag} = R{\rm ln}(2S + 1) = 17.29$ J/mol K for $S=7/2$. On the other hand, when using the \ymb\ lattice contribution, $S_{\rm {mag}}$ saturates to a value of 16.7 J/mol K which is slightly smaller than the theoretical prediction. The $S_{\rm {mag}}$ reaches to 94\% and 91\% of $R\ln(8)$ at $T_{\rm N}$ in these two calculations, respectively. The nonzero contributions to $C_{\rm mag}(T)$ and $S_{\rm mag}(T)$ above $T_{\rm N}$ arise from dynamic short-range magnetic order.  This feature has also been observed in other spin-7/2 Eu$^{+2}$ helical AFM systems~\cite{Sangeetha_EuCo2P2_2016, Sangeetha_EuCo2As2_2018, Sangeetha_EuNi2As2_2019}.

\begin{figure}
\includegraphics[width = 3.in]{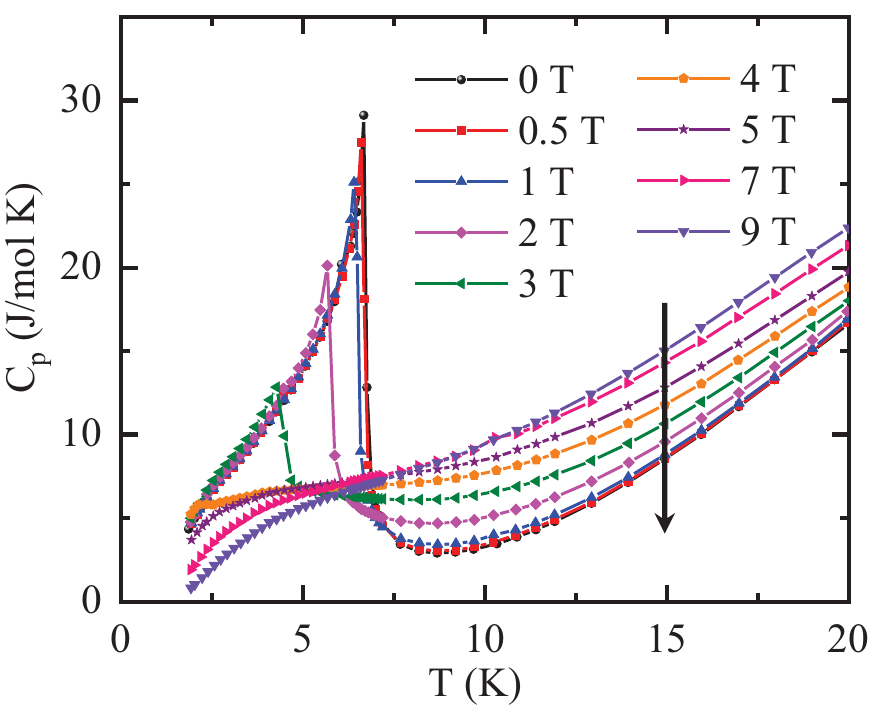}
\caption{Low-temperature heat capacity $C_{\rm p}(T)$ versus temperature~$T$ of \emb\ in different applied magnetic fields applied along the $c$~axis of the crystal as listed.  The arrow on the right crosses the data for increasing magnetic fields from 0 to 9~T\@.}
\label{Cp_T_Diff_field}
\end{figure}

\begin{figure}
\includegraphics[width = 3.in]{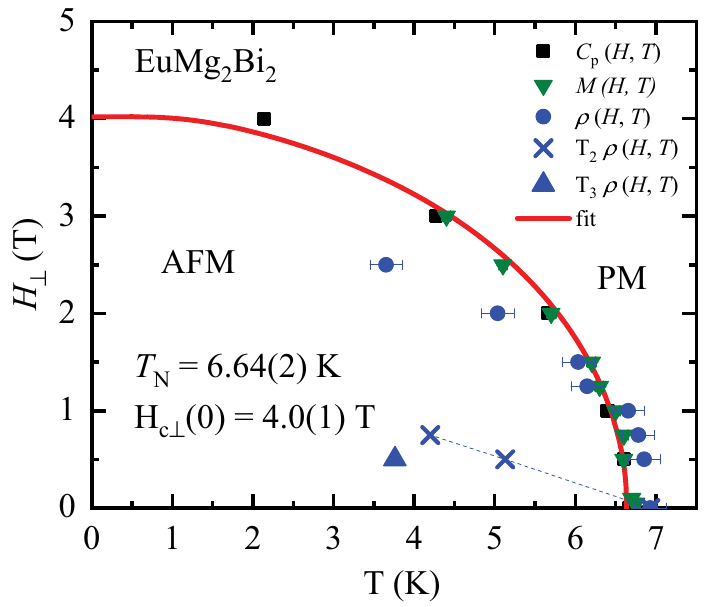}
\caption{Magnetic $H_\perp$-$T$ phase diagram constructed using the $C_{\rm p}(H_\perp, T)$ shown in Fig.~\ref{Cp_T_Diff_field}, where $H_\perp$ is defined to be the field along the $c$~axis, perpendicular to the $ab$ plane in which the moments order in zero field. The red curve is a fit to the critical-field $H_{\rm c\perp}(T)$ data by Eq.~(\ref{Eq:phase_diagram_MFT1}) for spin $S=7/2$ and the fitted parameters are listed in the figure. Data obtained from $M(H_{\rm c\perp}, T)$ and $\rho(H_{\rm c\perp}, T)$ measurements have also been plotted.  This fit separates the phase diagram into two regions which are the helical antiferromagnetic (AFM) and paramagnetic (PM) regions.  The three special points in the $\rho(T)$ data at low fields between 4.3 and 5.2~K ($T_2$, crosses) and at 3.7~K ($T_3$, up-triangle) are associated with phase transitions of unknown origin that were not detected in the other measurements.}
\label{Cp_H-T_Phase_diagram}
\end{figure}

\begin{figure}
\includegraphics[width = 3.in]{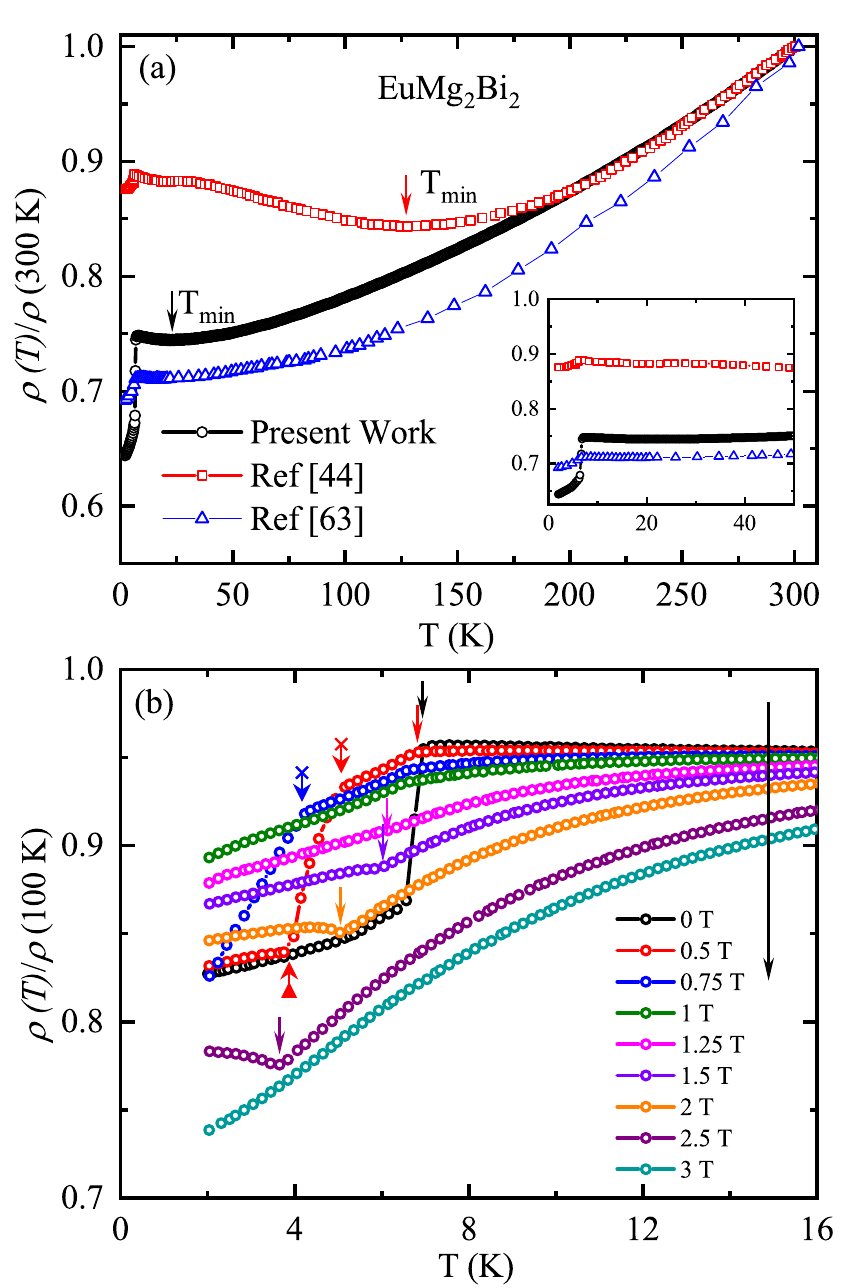}
\caption{(a) Zero-field $ab$-plane normalized resistivity ($\rho (T)/\rho (300 \rm K)$) of \emb\ in the temperature region 2--300~K\@. The data of the present work (black open circles) are compared with the earlier published data of the crystals of the same compound grown in a similar way (blue open triangles \cite{May2012}) and different way (red open squares \cite{Kabir2019}). The arrows indicate positions of resistivity minima at $T_{\rm min}$, absent in the samples with lowest residual resistivity above $T_N$.    Inset: expanded low-temperature region close to $T_{\rm N}$. (b)~Low-temperature region of the normalized resistivity $\rho (T)/\rho (100 \rm K)$ measured  for different applied magnetic fields with  $H \parallel c$. The long arrow points from low- to high-field data. Short arrows indicate features at the highest $T$ anomaly coinciding for low fields with $T_N$, cross-arrows (crosses in the phase diagram in Fig.~\ref{Cp_H-T_Phase_diagram}) and up-triangle arrow (up-triangle in Fig.~\ref{Cp_H-T_Phase_diagram}) point to additional features in $\rho(T)$ observed at temperatures below $T_{\rm N}$. }
\label{resistivity}
\end{figure}

The variation of $C_{\rm p}(T)$ with magnetic field~$H_\perp$ defined to be the field applied along the $c$ axis, perpendicular to the zero-field $ab$ ordering plane, is shown in Fig.~\ref{Cp_T_Diff_field}. Here we define the field along the $c$~axis as $H_\perp$ so as not to confuse the notation with the critical field $H_{\rm c}$ and the $c$-axis field $H_c$.  Due to the constraints of the PPMS used to measure $C_{\rm p}$, it was not possible to measure $C_{\rm p}(T)$ with $H\parallel ab$.  As seen from Fig.~\ref{Cp_T_Diff_field}, $T_{\rm N}$ shifts to lower temperature with increasing $H_\perp$ and also the jump in $C_{\rm p}$ at $T_{\rm N}$ decreases, where $T_{\rm N}$ is defined as the temperature of the peak in the heat capacity because of the contribution of dynamic short-range magnetic ordering to $C_{\rm mag}$ above $T_{\rm N}$\@.

From the data in Fig.~\ref{Cp_T_Diff_field}, one could plot $T_{\rm N}$ versus the field $H_\perp$  or the critical field $H_{\rm c\perp}$ versus~$T$\@.  We use the latter scaling because for a helical antiferromagnet,  $H_{c\perp}(t)$ is given theoretically by MFT as~\cite{Johnston2015}
\bea
\label{Eq:phase_diagram_MFT1}
H_{c\perp}(t) = H_{c\perp}(t=0)\bar{\mu}_0(t),
\eea
where $\bar{\mu}_0(t)$ is calculated using Eqs.~(\ref{Eqs:Cmag}). A fit of the $H_{c\perp}$ versus~$T$   data in Fig.~\ref{Cp_H-T_Phase_diagram} by Eq.~(\ref{Eq:phase_diagram_MFT1}) yields \mbox{$T_{\rm N} = 6.64(2)$~K} and $H_{c\perp}(T=0) = 4.0(1)$~T, as shown by the red curve in the figure.  This curve divides the phase diagram in the $(H_\perp,T)$ plane into helical AFM and PM regions as shown.


\section{\label{Sec:rho} Electrical Resistivity}

The temperature dependence of the $ab$-plane electrical resistivity measured in the temperature range 2 to 300 K in zero magnetic field is plotted in Fig.~\ref{resistivity}(a). The data are presented using a normalized $\rho (T)/\rho (300~ \rm K)$ scale. The resistivity shows a metallic decrease on cooling down to a shallow minimum at $T_{\rm min} \sim 23$~K, followed by a slight increase on further cooling.  A sharp resistivity decrease below $T = 6.75$~K (transition midpoint) indicates the onset of AFM ordering in the system. 

The observed $\rho (T)$ behavior is similar to that reported in similarly-grown crystals, blue triangles in Fig.~\ref{resistivity}(a)~\cite{May2012}, but notably different from that reported in the previous study~\cite{Kabir2019} in which a Sn-flux-grown single crystal was used for the measurements. The shallow minimum in $\rho (T)$ as observed in our study at about 23~K is not observed in similarly-grown samples with higher RRR and is observed at a much higher temperature of \mbox{$\sim 125$~K,} in samples with notably lower RRR\@. Using the value of resistivity above $T_N$ as a proxy for $\rho (0)$, the extrapolated  residual-resistivity ratio RRR~$\equiv \rho(300$~K)/$\rho(T_{\rm N})$ increases from $\approx 1.14$ in the Sn-flux-grown crystal, to $\approx 1.33$ in our crystals and to $\approx 1.39$ in the data of Ref.~\cite{May2012}.  Progressively the position of the resistivity minimum shifts to zero.  In previous studies of polycrystalline EuMg$_2$Bi$_2$~\cite{May2012} the minimum was much more pronounced and located at about 100~K\@.  The slight resistivity increase on cooling is accompanied by a temperature-independent Hall constant~\cite{May2012} arguing against an activated  character of charge transport below the minimum. The minimum was also observed in nonmagnetic CaMg$_2$Bi$_2$, which makes a possible contribution of the Kondo effect~\cite{Kondoeffect} unlikely.  We conclude that the most likely reason for the minimum is an onset of Anderson localization~\cite{Anderson} due to the strong effect of disorder in a low carrier-density metal. 

The decrease of resistivity below $T_{\rm N}$ in both previous studies is notably smaller than in our crystals, and is significantly less sharp. The origin of this discrepancy is unclear and deserve further study.   The resistivity decrease below $T_{\rm N}$ is governed by a decrease of magnetic entropy~\cite{Paglione2005} and the sharp feature may be suggestive of the first-order character of the transition.   The evolution of the temperature-dependent resistivity with $H\parallel c$ is shown in Fig.~\ref{resistivity}(b).  Of special note is the evolution between $H=0$~T~(black) and 0.5~T~(red). While the transition with onset at $T_{\rm N}=6.93$~K  in zero field is accompanied by a pretty sharp (full width of 0.4~K) resistivity decrease, the feature at $T_{\rm N} = 6.85$~K [which does not change much in temperature compared to the 0~T curve, as indicated by simple vertical black and red arrows in Fig.~\ref{resistivity}(b)], the decrease in  $\rho(T)$ at 0.5~T is quite small.   On further cooling $\rho(T)$ for 0.5~T reveals a second feature [we label it $T_2=5.1$~K and indicate with a cross arrow in Fig.~\ref{resistivity}(b)], below which the main resistivity decrease happens. A similar $T_2$ feature is found for the curve in 0.75~T field. The third feature  at $T_3=3.8$~K (up-triangle arrow) leads to flattening of the $\rho(T)$ in $H_c=0.5$~T for $T\to 0$~K\@.  The $T_3$ is not found  for 0.75~T in the temperature range studied.  A magnetic field of 1~T suppresses the $T_2$ feature, but the feature at $T_{\rm N}$ remains clearly discernible.  With further magnetic field increase to 1.25~T  the shape of the anomaly in $\rho(T) $ changes qualitatively, with initial  flattening and eventual increase of resistivity on cooling for $H_c=2.5$~T\@.  A monotonically decreasing $\rho(T)$ as found at 3~T suggests the suppression of $T_{\rm N}$ to below the temperature range studied.

We summarize the anomalies in $\rho(T,H)$ in the above phase diagram in Fig.~\ref{Cp_H-T_Phase_diagram}.  For fields below 1~T the feature in $\rho(T,H)$ at $T_{\rm N}$ (solid blue circles) is in good agreement with the heat capacity  (black solid squares)  and magnetization (green down-triangles) measurements.  For $H_\perp = 1.25$~T the position of the highest-temperature feature starts to go to zero notably faster than suggested by the magnetization and heat capacity determinations of $T_{\rm N}$. Note that the shape of $\rho(T)$ changes in the same field range.

The position of the $T_2$ feature (blue crosses in Fig.~\ref{Cp_H-T_Phase_diagram}) seems to be smoothly connected to the zero-field $T_{\rm N}$ (blue dashed line). The position of the $T_3$ feature (solid blue up-triangle) does not seem to be connected to any special point in the phase diagram.  

The sharpness of the feature at $T_{\rm N}$ in zero-field resistivity measurements suggest a first-order character of the transition, and its splitting into two second-order transitions with application of magnetic field.  However, this interpretation is not supported by either the $\chi(T)$ or $C_{\rm p}(T)$ data in Fig.~\ref{Chi_T_1kOe} and the inset of Fig.~\ref{Cp_T_zero_field}(a), respectively.

The fact that the low-temperature features at $T_2$ and $T_3$ are not observed in heat capacity and magnetization measurements may suggest that the features are due to lock-in transitions (transformations of helical order in magnetic field between incommensurate and commensurate on cooling), particularly difficult to observe in thermodynamic measurements due to the minute entropy changes involved~\cite{Jensen}.    

\section{\label{Sec:Summary} Summary}

In this work, we have investigated the detailed magnetic, thermal, and electronic transport properties of single crystals of the trigonal compound \emb, which has recently been reported to host multiple Dirac states. The magnetic susceptibility shows that the compound undergoes AFM ordering below $T_{\rm N} = 6.7 $ K associated with the Eu$^{+2}$ spins~7/2, as reported earlier.

The magnetic susceptibilities are found to be almost independent of temperature below $T_{\rm N}$ for both $H\parallel c$ and $H\parallel ab$, where the hexagonal setting of the trigonal structure has lattice parameters $a=b$ and~$c$.  According to molecular field theory (MFT), this behavior strongly suggests that the magnetic structure of \emb\ below $T_{\rm N}$ is a $c$-axis helix, where ferromagnetically-aligned moments in  $ab$ planes rotate in a $c$~axis helical structure by a turn angle of $\approx 120^\circ$ from plane to plane along the $c$~axis.  Another possible but less probable magnetic structure is a planar structure with in-plane nearest-neighbor Eu spins aligned at $\approx 120^\circ$ with respect to each other, with such planes stacked along the $c$~axis.  The latter structure is less likely because the calculated value of the net in-plane magnetic exchange interaction~$J_0$ derived from MFT is positive (FM), rather than negative (AFM) as would be expected for the latter structure, and because the Weiss temperature in the Curie-Weiss law is positive (FM-like) for fields in the $ab$~plane.

According to MFT, the $c~(\perp)$-axis magnetization $M_c$ of a $c$-axis helix is linear for applied fields from $H=0$ to the critical field $H_{{\rm c\perp}}$ at which the magnetization approaches the saturation magnetization and a second-order transition to the paramagnetic state occurs.  For fields applied in the $ab$ plane of a helix with a $120^\circ$ turn angle, $M_{ab}$ is also predicted to be linear from $H=0$ to the critical field $H_{{\rm c}\,ab}$ at $T=0$, even though a smooth crossover from a helix to fan phase occurs if the spins are confined to the $ab$~plane.  On the full scale of our $M(H)$ measurement field, these predictions are verified.  However, on closer examination, we find that $M_{ab}(H)$ shows a subtle nonlinearity at fields below about 500~Oe.  It would be interesting in future work to determine experimentally what change in the magnetic structure this nonlinearity is associated with.

The zero-field heat capacity $C_{\rm p}(T)$ measurement reveals a $\lambda$ anomaly at $T_{\rm N}$ that shifts to lower temperature with increasing $H$\@. The zero-field magnetic contribution $C_{\rm mag}(T)$ to $C_{\rm p}(T)$ obtained using two different background subtractions reveals the presence of short-range dynamic magnetic fluctuations both below and above $T_{\rm N}$ that contribute to the high-temperature limit of the magnetic entropy. This limit is close to the value expected for Eu spins $S=7/2$.

A sharp drop in the electrical resistivity $\rho (T)$ is observed on cooling below $T_{\rm N}$ in zero field. It is replaced by a two-stage resistivity decrease in the smallest applied magnetic fields.  This behavior is contrary to the previous studies and deserves further investigation.  This behavior is not reflected in the $\chi(T)$ or $C_{\rm p}(T)$ data, and to our knowledge has not been observed previously on cooling below $T_{\rm N}$ in any other antiferromagnetic material.  The drop is linear in temperature in nonzero $c$-axis magnetic fields with the temperature width of the drop increasing with increasing field.  A resistivity minimum above $T_{\rm N}$ was also observed in the $\rho (T)$ data at $\sim 23$~K, which is a significantly lower temperature than that reported earlier in Ref.~\cite{Kabir2019}. Interestingly, although only one magnetic transition at $T_{\rm N}$ is observed in our magnetic and heat capacity data, in addition to the above-noted feature at $T_{\rm N}$ the $\rho (T)$ data also reveal another distinct field-dependent anomaly in the magnetic-field range $0.5~{\rm T} \leq H \leq 0.75$~T for $T = 4.3$ to $5.2$~K\@.

On the basis of the magnetic, thermal, and electronic transport studies, a magnetic phase diagram in the $H$-$T$ plane for fields parallel to the $c$~axis was constructed that includes the antiferromagnetic and paramagnetic regions.  The phase boundary between these two phases is fitted satisfactorily by MFT\@.

It would be interesting to theoretically investigate the degree and manner to which one or more of our measured properties of single-crystalline \emb\ are  influenced or even caused by topological features of the band structure. Of particular interest is the origin of the rapid drop in the resistivity on cooling below~$T_{\rm N}$ in zero field.


\acknowledgments

We are grateful to P. P. Orth for helpful comments on the manuscript.  The authors thank J.~Jensen for useful discussion.  This research was supported by the U.S. Department of Energy, Office of Basic Energy Sciences, Division of Materials Sciences and Engineering.  Ames Laboratory is operated for the U.S. Department of Energy by Iowa State University under Contract No.~DE-AC02-07CH11358.

\end{document}